\documentclass[11pt,leqno]{article}

\usepackage{amsmath,amsthm}
\usepackage {latexsym}
\usepackage{amssymb}

\newcommand{\les}{\lesssim}

\newcommand{\bse}{\begin{equation}}

\newcommand{\bea}{\begin{eqnarray}}
\newcommand{\eea}{\end{eqnarray}}
\newcommand{\be}{\begin{equation}}
\newcommand{\ee}{\end{equation}}

\newcommand{\half}{\frac{1}{2}}

\newcommand{\R}{{\mathbb R}}

\newcommand{\calg}{\,{\mathfrak g}}

\newcommand{\Laplace}{\frac{\triangle}{2}}
\newcommand{\Lapl}{\frac12{\triangle}}

\newcommand{\trip}{|\!|\!|}
\newcommand{\la}{\langle}
\newcommand{\ra}{\rangle}

\def\pr{\partial}

\def\nn{\nonumber}
\def\calge1{\calg_{\vec{e_1}}}

\def\bm{\left( \begin{array}{cc}}
\def\endm{\end{array}\right)}

\def\Ran{{\rm Ran}}

\newtheorem{theorem}{Theorem}[section]
\newtheorem{lemma}[theorem]{Lemma}
\newtheorem{defi}[theorem]{Definition}
\newtheorem{cor}[theorem]{Corollary}
\newtheorem{prop}[theorem]{Proposition}

\theoremstyle{remark}
\newtheorem{remark}[theorem]{Remark}

\def\ga{\gamma}

\numberwithin{equation}{section}

\begin{document}

\title{ Fine Properties of Charge Transfer Models}
 \author{ Kaihua Cai
\thanks{The paper is for the author's candidacy of Ph.D. thesis at Caltech. The author
feels deeply grateful to his advisor Dr. Wilhelm Schlag. His
influence can be seen in every page. The author also wants to
thank Rowan Killip for his very helpful comments.}} \maketitle

\begin{abstract}
\noindent We prove $L^1 \rightarrow L^{\infty}$ estimates for
charge transfer Hamiltonians $H(t)$ in $\R^n$ for $n \geq 3$,
followed by a discussion on  $W^{\kappa,p'} \to W^{\kappa,p}$
estimates for the same model, where $2 < p < \infty $ and $\frac
1p + \frac 1{p'}=1$. Then, geometric methods are developed to
establish the time boundedness of the $H^\kappa$ norm for the
evolution of charge transfer operators and asymptotic completeness
of the Hamiltonian $H(t)$ in the $H^\kappa$ norm, where $\kappa$
is any positive integer.
\end{abstract}

\section{Introduction}
This paper is devoted to the study of the model corresponding to
the time-dependent charge transfer Hamiltonian
$$
H(t) = -\Lapl + \sum_{j=1}^{m} V_j (x - \vec v_j t)
$$
with rapidly decaying smooth stationary potentials $V_{j}(x)$ and
a set of mutually distinct constant velocities $\vec v_{j}$. First
we focus on the $L^1 \rightarrow L^{\infty}$ dispersive estimate
for the solutions of the time-dependent problem
$$
\frac 1i \partial_{t }\psi + H(t) \psi = 0
$$
associated with a charge transfer Hamiltonian $H(t)$. This kind of
dispersive estimate has been studied intensively during the past
twenty years. The starting point is the well-known $L^p$ estimates
for the free Schr\"odinger equation ($H_0=-\Lapl$) on $\R^n$:
$$ \| e^{itH_0} f \|_{L^p}\leq
C_p\, |t|^{-n(\frac{1}{2} - \frac{1}{p})}\|f\|_{L^{p'}},\quad
+\infty \geq p\geq 2,\,\,\, \frac{1}{p}+\frac{1}{p'}=1.
$$
 Writing the evolution as a convolution operator, the estimate is straight forward. One
application is that they imply the Strichartz estimates
$$
\| e^{itH_0} f \|_{L^q_t L^r_x}\leq C_q\|f\|_{L^2},\quad 2 \leq r,
q \leq \infty,\,\,\,\, \frac{n}{r} + \frac{2}{q}
=\frac{n}{2},\quad n\geq 3\qquad[ GV, KT]
$$
Such estimates play a fundamental role, among other things in the
theory of nonlinear dispersive equations.  The extension of such
theories motivated the efforts to establish the $L^p$ decay
estimates for the general {\it time independent} Schr\"odinger
operators of the type $H= -\Lapl + V(x)$. In this case there may
be bound states, i.e., $L^{2}$ eigenfunctions of $H$. Under the
evolution $e^{-itH}$, such bound states are merely multiplied by
oscillating factors and thus do not disperse. So we need to
project away any bound state and the estimates should take the
form
\begin{align}
&\|e^{-it H} P_{c}(H) \psi_{0}\|_{L^p}\leq C_p\, |t|^{-n(\frac{1}{2} -
\frac{1}{p})}\|\psi_{0}\|_{L^{p'}}\quad p\ge 2,\,\,\,
\frac{1}{p}+\frac{1}{p'}=1,\label{eq:inddec}\\
&\| e^{-itH} P_{c}(H)f \|_{L^q_t L^r_x}\leq C_q\|f\|_{L^2} \text{ for }
2 \leq q
\leq \infty,\,\, \frac{n}{r} + \frac{2}{q} =\frac{n}{2},\quad n\geq
3\label{eq:indStr}
\end{align}
where $P_{c} (H)$ is the projection onto the continuous part of
the spectrum of the self-adjoint operator $H$.

Before \eqref{eq:inddec} and \eqref{eq:indStr} were established,
Rauch~\cite{R}, Jensen and Kato~\cite{JK}, and Jensen~\cite{J}
proved the dispersion of $ e^{-itH} P_{c}(H)$ in weighted $L^2$
spaces. Rauch required exponential decay of the potential, whereas
Jensen and Kato assumed polynomial decay of a certain rate.
Because the $L^2$ norm is preserved by the evolution, we need only
prove \eqref{eq:inddec} for the case where $p=\infty$ by
interpolation. There are several approaches to the proof of
\eqref{eq:inddec}. In \cite{Ya1}, Yajima proved that the wave
operators are bounded on the Sobolev spaces $W^{\kappa, p}$ and
\eqref{eq:inddec} is a consequence of the intertwining property of
the wave operators. In \cite{JSS}, the evolution operator is
expanded by Duhamel's formula and its cancellation property was
explored. Their proof splits into two parts: a ``high energy"
estimate, which holds for all potentials, and a ``low energy"
estimate where some spectral property of $H$ is assumed. This
approach is generalized in \cite{RSS1} to prove a ``weak version"
of the dispersive estimate of $H(t)$ (see inequality
\eqref{eq:oldmain}). The first goal of this paper is to extend
this idea further and prove the dispersive estimate of $H(t)$.
Recently, M. Goldberg and W. Schlag (\cite{GS}) proved
\eqref{eq:inddec} with much less restrictive conditions on the
potential in $\R^1$ and $\R^3$. Their method is expected to give
\eqref{eq:inddec} in all dimensions.

We proceed by defining the charge transfer model and specifying
our basic assumption.

\begin{defi}
\label{def:chargetrans} By a {\em charge transfer model} we mean a
Schr\"odinger equation \bea & \frac{1}{i} \partial_t \psi - \Lapl
\psi + \sum^m_{\kappa =1}
V_\kappa(x - \vec v_\kappa t) \psi = 0 \label{eq:transfer} \\
& \psi |_{t=0} = \psi_0,  x\in \R^n, \nn \eea where $\vec
v_\kappa$ are distinct vectors in $\R^n$, $n\ge 3$, and the real
potentials $V_\kappa$ are such that for every $1\le\kappa\le m$,
\begin{enumerate}
\item $V_\kappa$ is time independent and has compact support (or
fast decay), $V_{\kappa}, \nabla V_\kappa \in L^\infty$.

\item $0$ is neither an eigenvalue nor a resonance of the
operators
$$
H_\kappa = - \Lapl + V_\kappa(x).$$
\end{enumerate}
\end{defi}

Recall that $\psi$ is a resonance if it is a distributional
solution of the equation $H_\kappa \psi =0$ which belongs to the
space $L^2(\langle x \rangle^{-\sigma}dx)$ for any $\sigma
>\frac 12$, but not for $\sigma =0$.

The conditions in the above definition is always assumed when we
prove and apply the dispersive estimates, i.e.
Theorem~\ref{thm:main} and Theorem~\ref{thm:main2}. The conditions
are not for optimal, but for convenience. This definition is
standard, see~\cite{Gr}, \cite{Ya2}. The Schr\"odinger group
$e^{-itH_\kappa}$ is known to satisfy the decay estimates (see
Journ\'e, Soffer, Sogge~\cite{JSS} and Yajima~\cite{Ya1})
\begin{equation}
\| e^{-itH_\kappa} P_c (H_\kappa) \psi_0 \|_{L^\infty} \lesssim
|t|^{-\frac n2} \|\psi_0\|_{L^1} \label{eq:nodecay}
\end{equation}
for $n\geq 3 $ under various conditions on the potential. Here
 $P_c(H_\kappa)$ is the spectral projection onto the continuous spectrum of $H_\kappa$ and
$\lesssim$ denotes bounds involving multiplicative constants
independent of $\psi_0$ and $t $. For $n=3$, \cite{GS} proved
\eqref{eq:nodecay} under the assumption that $|V_\kappa(x)| \leq C
(1+|x|)^{-\beta} $, for some $ \beta >3$. For $n>3$,
\eqref{eq:nodecay} holds (\cite{JSS}) under the additional
assumption: ${\mathcal{F}} ( V_\kappa ) \in L^1$.
Yajima~\cite{Ya1} proved \eqref{eq:nodecay} with slightly weaker
conditions than \cite{JSS}. We shall assume that ${\mathcal{F}}
(V_{\kappa}) \in L^1 $ to guarantee the estimate
\eqref{eq:nodecay}, except in Section 5.

To establish similar dispersive estimates for time dependent
Schr\"odinger equations is more involved. Heuristically, we can't
project away the bounded states as they are moving in different
directions. Rodnianski and Schlag~\cite{RS} proved dispersive
estimate for small time dependent potentials. In this paper, we
will focus on the charge transfer model.

An indispensable tool in the study of the charge transfer model
are the Galilean transforms
\begin{equation} \label{eq:gal}
\calg_{\vec{v},y}(t)= e^{-i\frac{|\vec v|^2}{2} t} e^{-i{x\cdot\vec
v}} e^{i(y+t\vec v)\cdot\vec p},
\end{equation}
cf.~\cite{Gr}, where $\vec p=-i\vec \nabla$. Under
$\calg_{\vec{v},y}(t)$, the Schr\"odinger equation transforms as
follows:
\be
 \calg_{\vec v,y}(t) e^{it\Laplace} = e^{it\Laplace} \calg_{\vec
v,y}(0) \label{eq:galilei} \ee
 and moreover, with $H=-\Lapl + V$,
\begin{equation}
\label{eq:invg}  \psi(t) := \calg_{\vec v,y}(t)^{-1} e^{-itH}
\calg_{\vec v,y}(0) \phi_0, ,\qquad \calg_{\vec v,y}(t)^{-1} =
e^{-i{y\cdot \vec v}}\calg_{-\vec v,-y}(t)
\end{equation}
solves \bea & \frac{1}{i} \partial_t \psi - \Lapl \psi + V(\cdot -
t\vec v -y) \psi = 0 \label{eq:shifted} \\
& \psi |_{t=0} = \phi_0. \nn \eea

Since in our case always $y=0$, we set $\calg_{\vec v}(t):=
\calg_{\vec v,0}(t)$. Note that the transformations $\calg_{\vec
v,y}(t)$ are isometries on all $L^p$ spaces and
$\calge1(t)^{-1}=\calg_{-\vec{e_1}}(t)$ because
of~\eqref{eq:invg}.  In the following, we shall assume that the
number of potentials is $m = 2$ and that the velocities are $\vec
v_1 = 0, \vec v_2 = (1,0, \dots 0) = \vec e_1$. The arguments
generalize easily to $m \geq 3 $.

We now introduce the appropriate analog to project away bounded
states for the problem \bea && \frac{1}{i}
\partial_t \psi - \Lapl \psi + V_1 \psi +V_2
(\cdot -t\vec{e_1}) \psi = 0 \label{eq:2pot} \\
&& \psi|_{t=0} = \psi_0 \nn \eea with compactly supported
potentials $V_1, V_2$. Let $u_1, \ldots, u_m$ and $w_1, \ldots,
w_\ell$ be the normalized bound states of~$H_1$ and $H_2$
corresponding to the negative eigenvalues $\lambda_1, \dots,
\lambda_m$ and $\mu_1, \dots, \mu_\ell$, respectively (recall that
we are assuming that $0$ is not an eigenvalue). We denote by
$P_b(H_1)$ and $P_b(H_2)$ the corresponding projections onto the
bound states of $H_1$ and $H_2$, respectively, and let
$P_c(H_\kappa)=Id-P_b(H_\kappa)$, $\kappa=1,2$. The projections
$P_b(H_{1,2})$ have the form
$$
P_b(H_1) =\sum^m_{i=1} \langle \cdot, u_i\rangle u_i, \qquad
P_b(H_2) = \sum^\ell_{j=1} \langle \cdot, w_j\rangle w_j.
$$
The following orthogonality condition in the context of the charge
transfer Hamiltonian~\eqref{eq:2pot} was introduced in
\cite{RSS1}.

\begin{defi}
\label{def:asymp}  Let $U(t) \psi_0 = \psi(t, x)$ be the solutions
of~\eqref{eq:2pot}. We say that $\psi_0$ (or also $\psi(t,\cdot)$)
is {\em asymptotically orthogonal} to the bound states of $H_1$
and $H_2$ if
\begin{equation}
\label{eq:scat}
\|P_b(H_1)U(t) \psi_0 \|_{L^2} + \| P_b(H_2,t) U(t)\psi_0 \|_{L^2}\to
0\text{ as }t\to +\infty.
\end{equation}
Here
\begin{equation}
\label{eq:Proj2}
P_b(H_2,t) := \calg_{\vec{-e_1}}(t) P_b(H_2)\, \calge1(t)
\end{equation}
for all times $t$.
\end{defi}

\begin{remark} From Corollary~\ref{cor:shorttime}, $\| U(t) \psi_0 \|_{L^p} \leq  C_t
\|\psi_0 \|_{L^{p'}}$, we know that $U(t) \psi_0 \in L^p$ is
well-defined for $\psi_0 \in L^{p'}$. As the bound states $u_i,
w_j$ are exponentially decaying at infinity,
Definition~\ref{def:asymp} makes sense for any initial data
$\psi_0 \in L^{p'} $ for $p' \in [1,2]$.
\end{remark}

\begin{remark} Clearly, $P_b(H_2,t)$ is again an orthogonal
projection for every~$t$. It gives the projection onto the bound
states of $H_2$ that have been translated to the position of the
potential $V_2(\cdot-t\vec{e_1})$. Equivalently, one can think of
it as translating the solution of~\eqref{eq:2pot} from that
position to the origin, projecting onto the bound states of $H_2$,
and then translating back.
\end{remark}

\begin{remark}
From Proposition 3.1 of \cite{RSS1}, the decay rate of
\eqref{eq:scat} is actually exponential. More precisely, the
following holds:

\begin{equation}
\label{eq:scat1} \|P_b(H_1)U(t) \psi_0 \|_{L^2} + \| P_b(H_2,t)
U(t)\psi_0 \|_{L^2} \lesssim e^{-\alpha t} \|\psi_0 \|_{L^2},
\end{equation}
for some $\alpha >0$.
\end{remark}

\begin{remark}
\label{rem:graf} It is clear that all $\psi_0$ that
satisfy~\eqref{eq:scat} form a closed subspace. This subspace
coincides with the space of {\em scattering states} for the charge
transfer problem. The latter is well-defined by Graf's asymptotic
completeness result~\cite{Gr}.
\end{remark}

We can only expect the dispersive estimate for \eqref{eq:2pot} for
the initial data satisfying Definition~\ref{def:asymp}, just as we
have to project away the bound states for \eqref{eq:nodecay}.
Rodnianski, Schlag, Soffer~\cite{RSS1} established the following
estimate

\begin{equation}
  \| U(t) \psi_0 \|_{L^2 + L^\infty} \lesssim
\langle t\rangle^{-\frac n2}\|\psi_0\|_{L^1\cap L^2}.
\label{eq:oldmain}
\end{equation}
with initial data $\psi_0 \in L^1 \cap L^2$ satisfying
\eqref{eq:scat}, where $U(t)$ is the evolution of the charge
transfer model and $\langle t\rangle = (1+t^2)^{\half}$. By
definition, $ \| f \|_{L^2 + L^\infty} := {\inf}_{f=h+g} (\| h
\|_{L^2} + \|g\|_{L^\infty}) $ and $\|f\|_{L^1\cap
L^2}=\|f\|_{L^1}+\|f\|_{L^2}$. \eqref{eq:oldmain} has important
application to the asymptotic stability and asymptotic
completeness for the small perturbation of non-colliding solitons
for NLS (\cite{RSS2}).

\cite{RSS1} decomposes the evolution into different channels
according to each potential. Every channel splits into a large
velocity part and a low velocity part. For the large velocity
part, they employed Kato's smoothing estimate; for the low
velocity part, a propagation estimate is used. In this paper, we
will combine the methods from \cite{JSS} and \cite{RSS1} and
obtain the following:

\begin{theorem}
\label{thm:main} Consider the charge transfer model as in
Definition~\ref{def:chargetrans} with two potentials,
cf.~\eqref{eq:2pot}. Assume $\widehat{V_1}, \widehat{V_2} \in
L^1(\R^n)$. Let $U(t)$ denote the propagator of \eqref{eq:2pot}.
Then for any initial data $\psi_0 \in L^1$, which is
asymptotically orthogonal to the bound states of $H_1$ and $H_2$
in the sense of Definition~\ref{def:asymp}, one has the decay
estimates
\begin{equation}
\| U(t) \psi_0 \|_{L^\infty} \lesssim |t|^{-\frac n2}
\|\psi_0\|_{L^1}. \label{eq:main}
\end{equation}
An analogous statement holds for any number of potentials, i.e., with
arbitrary $m$ in~\eqref{eq:transfer}.
\end{theorem}

Inspection of the argument in the following sections shows that it
applies to exponentially decaying potentials, say. But also
sufficiently fast power decay at infinity is allowed. We shall
prove~\eqref{eq:main} by means of a bootstrap argument. More
precisely, we prove that the {\em bootstrap assumption}
\begin{equation}
\| U(t) \psi_0\|_{L^\infty }\leq C_0  |t|^{-\frac n2}
\|\psi_0\|_{L^1} \text{\ \ for all\ \ }0\le t\le T\label{eq:boot}
\end{equation}
implies that
\begin{equation}
\|U(t) \psi_0\|_{L^\infty} \leq \frac{C_0}{2} |t|^{-\frac n2}
\|\psi_0\|_{L^1} \text{\ \ for all\ \ }0\le t\le T.
\label{eq:boot/2}
\end{equation}
Here $T$ is any given fixed large constant and \eqref{eq:boot}
holds for $C_0$, some sufficiently large constant because of
Corollary~\ref{cor:shorttime}. $C_0$ may depend on $T$ at the
beginning. The above implication
\eqref{eq:boot}~$\Longrightarrow$~\eqref{eq:boot/2} holds as long
as $C_0$ is larger than some universal constant independent of the
time $T$. Thus iterating
\eqref{eq:boot}~$\Longrightarrow$~\eqref{eq:boot/2} then yields a
constant that does not depend on~$T$. The theorem follows by
letting $T \to +\infty$.

As the $L^2$ norm of $U(t)\psi_0$ remains constant, by
interpolation, the following holds:

\begin{equation}
\| U(t) \psi_0 \|_{L^p} \lesssim C_p |t|^{-n(\frac 12- \frac 1p)}
\|\psi_0\|_{L^{p'}} \quad p\ge 2,\,\,\,
\frac{1}{p}+\frac{1}{p'}=1.
\end{equation}

Our next theorem is about the decay estimates of $\partial^\alpha
U(t)\psi_0$, where $\alpha=(\alpha_1,\cdots,\alpha_n)$ is an
n-tuple of nonnegative integers and $\partial^\alpha = \frac
{\partial^\alpha}{\partial x_1^{\alpha_1} \cdots \partial
x_n^{\alpha_n}}$. We write $|\alpha|=\alpha_1+ \cdots +\alpha_n $.

\begin{theorem}
\label{thm:main2}  Let $U(t)$ denote the propagator of the
equation~\eqref{eq:2pot}. Assume \eqref{eq:nodecay} holds for
$H_1$ and $H_2$. Let $V_j \in C_0^{\kappa +1} $ where $\kappa$ is
a positive integer and $j=1,2$. Moreover, assume that for $\forall
|\beta| \leq \kappa$ and $j=1,2$, $\widehat{\partial ^\beta V_j}
\in L^1(\R^n)$. Then for any initial data $\psi_0 \in
W^{\kappa,p'}$, which is asymptotically orthogonal to the bound
states of $H_j \,\, (j=1, 2)$ in the sense of
Definition~\ref{def:asymp}, one has the decay estimates

\begin{equation}
\| U(t) \psi_0 \|_{W^{\kappa,p}} \lesssim |t|^{-n (\frac 12-\frac
1p)} \|\psi_0\|_{W^{\kappa,p'}}. \label{eq:main2}
\end{equation}
\noindent  where $ 2 \leq p < \infty $ and $\frac 1p +\frac
1{p'}=1$.
\end{theorem}

\begin{remark} It suffices to prove Theorem~\ref{thm:main2} for $p > \frac {2n}{n-2}
$, because interpolating with Theorem~\ref{thm:main3}, which holds
under the assumption of Theorem~\ref{thm:main2}, we derive
Theorem~\cite{thm:main2} for any $p \in [2, +\infty]$.  $p > \frac
{2n}{n-2} $ guarantees that $\int _1^{\infty}|t|^{-n (\frac
12-\frac 1p)} < \infty$. We need to exclude the case $p =\infty $,
since part of our proof relies on singular integrals and we do not
know whether or not \eqref{eq:main2} holds for $p=\infty$.
\end{remark}

The second part of this paper is motivated by Graf~\cite{Gr}. Graf
proved energy boundedness for $U(t,s)$ by a geometric method,
where $U(t,s)$ is the solution operator corresponding to the
time-dependent Schr\"odinger equation

\bea &&\frac 1i \partial_{t}\psi -\Laplace \psi + \sum^m_{j =1}
V_j(x - \vec v_j t) \psi =0,
\label{eq:timeSc'}\\
&&\psi|_{t=s} =\psi_{0},\nn \eea i.e., $\psi(t,\cdot) =
U(t,s)\psi_{0}$. \cite{Gr} proved that $\|U(t,s) \psi_0\|_{H^1}$
is bounded as $t \to \infty$ provided that the initial data
$\psi_0 \in H^1(\R^n)$, $n \geq 1$. For the higher degree Sobolev
norms, J. Bourgain~\cite{Bo} proved the following for the general
time dependent Hamiltonian $H(t)$:

Suppose the time dependent potential $V(x,t) $ is bounded, real
and $\sup_{t \in \R} |V(x,t)| $ is compactly supported. Moreover,
for any n-tuple $ \alpha$

$$ \sup_{t\in \R} \|D^{\alpha}_x V(t) \|_\infty < C_\alpha. $$

 Then for $\forall \epsilon >0$ and $\kappa >0$,

 \be \label{eq:bourgain} \|U(t,0) \psi_0\|_{H^\kappa} \leq
C_{\epsilon,\kappa} |t|^{\epsilon} \|\psi_0\|_{H^\kappa} \qquad
\qquad \text{for all} t. \ee

An example (\cite{Bo}) is given to show that we can not remove the
$|t|^{\epsilon}$ growth for general time dependent potentials.
From this paper, it is shown that \eqref{eq:bourgain} does hold
with $\epsilon =0 $ for the case of the charge transfer
Hamiltonian. More precisely, in Section 4, the time-boundedness of
$\|U(t,s) \psi_0\|_{H^\kappa}$ for Charge Transfer Models is
established by the same geometric method as in \cite{Gr} for any
real number $\kappa$. We write $\lceil x \rceil $ as the least
integer no less than $x$. The precise statement is as follows:

\begin{theorem} \label{thm:main3}
Let $U(t,s)$ be the evolution operator for \eqref{eq:timeSc'}, and
let $\kappa \in \R $ and the dimension $n \geq 1$. Furthermore,
suppose $V_j \in C_0^{\lceil |\kappa| \rceil }(\R^n)$,
$(j=1,2,\cdots, m)$, i.e. $V_j$ has derivatives up to degree
$\lceil |\kappa| \rceil$, which are all continuous and compactly
supported. Then for $\forall t,s \in \R$
$$\|U(t,s)\psi_0\|_{H^\kappa} \leq C_\kappa \|\psi_0\|_{H^\kappa},$$
where $C_\kappa$ depends on $\kappa$ and the potentials $V_j$.

\end{theorem}

\begin{remark}  By interpolation, it clearly suffices to consider the case where $\kappa$
is an  integer. By duality, it suffices to prove the case where
$\kappa$ is a positive integer. Indeed, assuming $\kappa <0$, due
to the fact that $U(t,s)$ is unitary on $L^2 (\R^n)$, we have
\begin{align*} &
\|U(t,s)\psi_0\|_{H^\kappa}= \sup_{\| \phi \|_{H^{-\kappa}}=1}
\langle U(t,s)\psi_0 , \phi \rangle_{ L^2} \\
= & \sup_{\| \phi \|_{H^{-\kappa}}=1} \langle \psi_0 , U(s,t)\phi
\rangle_{L^2} \leq C_{-\kappa} \| \psi_0\|_{H^\kappa}.
\end{align*}

No assumption is made on the spectra of the subsystems $H_j$. The
assumption of compact support of $V_j$ is for convenience only and
the proof works for sufficiently fast polynomial decay at infinity
without essential change (\cite{Gr}). Suppose all assumptions of
both Theorem~\ref{thm:main2} and Theorem~\ref{thm:main3} hold,
then by interpolation, the estimate~\eqref{eq:main2} holds for $2
\leq p < \infty $.
\end{remark}
\begin{remark}
It follows from Duhamel's formula and Gronwall's inequality, that

\be \|U(t,s)\psi_0\|_{H^\kappa} \leq C(I) \|\psi_0\|_{H^\kappa}
\quad t,s \in I, \label{eq:compact} \ee for any compact interval
$I$. Therefore, it suffices to prove Theorem~\ref{thm:main3} when
$|t|$ or $|s|$ is large.
\end{remark}

As an important consequence, we apply Theorem~\ref{thm:main2} and
Theorem~\ref{thm:main3} to obtain the following asymptotic
completeness for the charge transfer model in the $H^{\kappa}$
sense:

\begin{theorem}
\label{thm:main4} Let $u_1,\ldots, u_m$ and $w_{1},\ldots, w_\ell$
be the eigenfunctions of $H_1 =- \Laplace + V_{1}(x)$ and
$H_2=-\Laplace + V_{2}(x)$, respectively, corresponding to the
negative eigenvalues $\lambda_1, \cdots, \lambda_m$ and $\mu_1,
\cdots, \mu_\ell$. Assume that $V_j \in C_0^{n+2\kappa+2}(\R^n)$,
$(n \geq 3,\,\, j=1,2)$, and that $0$ is neither an eigenvalue nor
a resonance of $H_1,H_2$, where $\kappa$ is a nonnegative integer.
Then for any initial data $\psi_0 \in H^2$, the solution
$U(t)\psi_{0}$ of the charge transfer problem \eqref{eq:2pot} can
be written in the form
\begin{equation}
\nn U(t) \psi_0 = \sum^m_{r=1} A_r e^{-i\lambda_rt} u_r +
\sum^\ell_{k=1}B_k e^{-i\mu_k t} \calg_{-\vec e_{1}} (t) w_k +
e^{-it\Laplace} \phi_{0} + \mathcal {R}(t),
\end{equation}
for some choice of the constants $A_{r}, B_{k}$ and the function
$\phi_{0} \in H^\kappa$. The remainder term ${\mathcal {R}}(t)$
satisfies the estimate
\begin{equation}
\nn \|\mathcal {R}(t)\|_{H^{\kappa}}\longrightarrow 0, \quad
\text{as}\,\, \,t\to \infty .
\end{equation}
\end{theorem}

\begin{remark}
The above theorem holds for $m$ potentials. We are not aiming to
give the optimal regularity condition on the potentials. The
theorem is equivalent to claiming that $ H^\kappa(\R^n)$ is the
sum of the ranges of the wave operators $\Omega^-_{l}$,
$(l=0,1,2)$ defined in Section 6.1. \cite{Gr} proved that the
ranges of the wave operators are orthogonal to each other in the
$L^2$ sense. Therefore, $ H^\kappa(\R^n)$ again is a direct sum of
$ \Omega^-_{l}(H^\kappa )$.
\end{remark}

\section{Cancellation Lemma}
\label{sec:cancel}

The first ingredient of our proof is the notion of cancellation.
In this section, $U(t)$ will denote the evolution operator of
\eqref{eq:2pot} or \eqref{eq:transfer}. It is clear from their
proofs that the following lemmas also hold for general time
dependent Hamiltonian $H_0 + V(t)$.

\begin{lemma}
\label{lemma:canc1}
\begin{equation}
\sup_{-\infty < t < \infty}\|e^{it\Delta}Ve^{-it\Delta}\|_{p\to p}
\leq \| \widehat{V} \|_1 ,  \label{eq:canc1}
\end{equation}
\end{lemma}
\noindent where $p \in [1,\infty]$ and $\|\cdot \|_{p \to p}$
means the operator norm from $L^p$ to $L^p$. For the proof of the
lemma, just notice that equation~\eqref{eq:galilei} implies
$[e^{it\Delta}e^{i\zeta x }e^{-it\Delta}f]
(x)=\calg_{-\zeta}(2t)f(x) =e^{-it|\zeta|^2} e^{ix\zeta} f(x-2
\zeta t )$.

\begin{lemma} \label{lemma:canc2}
Suppose  $t,s \in \R$, then we have the following:

\begin{equation}
\sup_{r \in \R} \|e^{-i(t-s)H_0}V(r)U(s)\|_{1\to \infty} < |t|^
{-\frac n2}C M e^{M|s|},
  \label{eq:canc2}
\end{equation}
\noindent where $M=\max_{r\in \R} \| \widehat{V(r)}\|_1 < \infty
.$

\end{lemma}
\begin{proof} Let's write $ \Psi (t,s):=\sup_{r \in \R} \|e^{-i(t-s)H_0}V(r)U(s)\|_{1\to \infty}
$. Without loss of generality, we suppose that $s>0$. By Duhamel's
formula,
$$ e^{-i(t-s)H_0}V(r)U(s)= e^{-i(t-s)H_0}V(r)\{e^{-isH_0}-i \int^s_0
e^{-i(s- \tau) H_0}V(\tau)U(\tau) d \tau \},$$
it follows that

\begin{align*} & \|e^{-i(t-s)H_0}V(r)U(s)\|_{1\to \infty} \\
 & \leq \|e^{-i(t-s)H_0}V(r)e^{-isH_0}\|_{1\to \infty} +
  \int^s_0 \|e^{-i(t-s)H_0}V(r) e^{-i(s-\tau) H_0}V(\tau)U(\tau)\|_{1\to \infty}
  d\tau \\
 & \leq C \|\widehat{V}(r) \|_1 |t|^{-\frac n2} + \|\widehat{V}(r)\|_1 \int^s_0
 \|e^{-i(t-\tau)H_0}V(\tau)U(\tau)\|_{1\to \infty} d \tau \\
 &\leq CM |t|^{-\frac n2} + M \int^s_0 \Psi (t,\tau) d \tau.
\end{align*}
Taking the supremum over $r$, we get $\Psi(t,s)\leq CM |t|^{-\frac
n2}+ M \int^s_0 \Psi (t,\tau)d \tau $. By Gronwall's inequality,
$$\Psi(t,s)\leq CM |t|^{-\frac n2} e^{Ms}. $$
\end{proof}

Note that the lemma still holds with other constants $C$ and $M$
on the right-hand side if we replace $V(r)$ with $V_j(r)$ or
replace $U(s)$ with another evolution, say $e^{-isH_j}$. Another
observation is that the lemma can be generalized to the following
by the same proof:

\begin{equation}
\sup_{r \in \R} \|e^{-i(t-s)H_0}V(r) U(s)\psi_0\|_p \lesssim
|t|^{-\gamma} M e^{Ms}\| \psi_0 \|_{p'} \label{eq:canc2'}
\end{equation}
\noindent where $\gamma = n(\frac 12 - \frac 1p)$ and $2 \leq p
\leq \infty$, $\frac 1p + \frac 1{p'}=1 $. This will be useful in
Section 4.

\begin{cor} \label{cor:shorttime}
Suppose $U(t)$ is the evolution operator of \eqref{eq:2pot} or
\eqref{eq:transfer}. Assume $t>0$, then
\begin{equation}
\|U(t)\|_{p'\to p} \lesssim t^{-n(\frac 12 - \frac 1p)} e^{Mt}
\quad \frac 1p + \frac 1{p'}=1, \, 2 \leq p \leq \infty
\label{eq:shorttime}
\end{equation}
\end{cor}

\begin{proof}
By Duhamel's formula, $ U(t)=e^{-itH_0}-i \int^t_0 e^{-i(t- \tau)
H_0}V(\tau)U(\tau) d\tau $. Write $\gamma = n(\frac 12 - \frac
1p)$, then by Lemma~\ref{lemma:canc2}, we have

$$\|U(t)\|_{p'\to p}
\leq C t^{-\gamma} + \int^t_0 \Psi(t,\tau)d \tau \leq C
t^{-\gamma} + \int^t_0 C t^{-\gamma} M e^{M\tau} d \tau \leq C
t^{-\gamma} e^{Mt} $$
\end{proof}

\noindent From the corollary, the bootstrap
assumption~\eqref{eq:boot} holds for any time $T$ if we take $C_0=
C e^{MT}$.

\begin{lemma} \label{lemma:canc3}
Suppose $m \geq 1$ and $\epsilon >0$. If $u_1, u_2,\ldots,u_m$ are
either all positive or all negative, satisfying
$|\sum_{j=1}^{m}u_j|> \epsilon$, then there exists a constant
$C=C(m,\epsilon)$ such that

\begin{align} \label{eq:canc3}
& \|\prod_{j=1}^{m-1} (e^{iu_j H_0}V(s_j)) e^{iu_m H_0}\|_{1\to
\infty} \leq C M^{m-1} \prod_{j=1}^m \langle u_j \rangle^{-\frac n2} \\
& \|\prod_{j=1}^{m-1} (e^{iu_j H_0}V(s_j)) U(u_m)\|_{1\to \infty}
\leq  C M^{m-1} \prod_{j=1}^m \langle u_j \rangle^{-\frac n2} e^{M
u_m} \label{eq:canc}
\end{align}

\noindent where $s_j$ is any real number and
$M=\sup_{s \in R}(\|V(s)\|_1 + \|\widehat{V}(s)\|_1)$.
\end{lemma}

\begin{proof}
The first inequality is from \cite{JSS}. Assume that $u_1,
u_2,\ldots,u_m$ are all positive without loss of generality. We
apply the dispersive estimate for $e^{iu_j H_0}$ repeatedly and
the left-hand side is dominated by $ C M^{m-1} \prod_{j=1}^m
u_j^{-\frac n2}$, which is dominated by the right-hand side up to
a constant, provided each $u_j > \epsilon$. If some $u_j \leq
\epsilon$, it is inefficient to use a dispersive estimate for
$e^{iu_j H_0}$. Instead, we apply the cancellation
lemma~\ref{lemma:canc1} and obtain

$$ e^{iu_j H_0} V(s_j) e^{iu_{j+1}
H_0}= \int e^{iu_j H_0} e^{ix \zeta} e^{-iu_j H_0}
\widehat{V(s_j)}(\zeta)\, d\zeta e^{i(u_{j+1}+u_j) H_0}.
$$
where $ e^{iu_j H_0} e^{ix \zeta} e^{-iu_j H_0}$ is the Galilean
transform $ \calg_{-\zeta}(-u_j)$ according to \eqref{eq:galilei}.
If again $u_j + u_{j+1} < \epsilon $, we can repeat this procedure
until $u_{j-l}+ \cdots + u_j + \cdots + u_{j+k} > \epsilon $ which
always happens because $|\sum_{j=1}^{m}u_j|> \epsilon$. Then we
apply the dispersive estimate to obtain the inequality.

We sketch the proof of the second equation. When $m=1$, it is just
\eqref{eq:shorttime} provided that $u_m > \epsilon $. When $m=2$,
if $u_1 > \frac {\epsilon}2$ and $u_2 > \frac{\epsilon}2$,

\begin{align*} & \|(e^{iu_1 H_0}V(s_1)) U(u_2)\|_{1\to \infty}
\lesssim |u_1|^{-\frac n2} \|V(s_1) U(u_2)\|_{1\to 1} \\
& \lesssim |u_1|^{-\frac n2} \|U(u_2)\|_{1\to \infty} \\
& \lesssim |u_1|^{-\frac n2} |u_2|^{-\frac n2} e^{Mu_2} \lesssim
\langle u_1 \rangle^{-\frac n2} \langle u_2 \rangle^{-\frac
n2}e^{Mu_2} \end{align*}

If $u_1 \leq \frac {\epsilon}2$ or $u_2 \leq \frac{\epsilon}2$, we
apply  Lemma~\ref{lemma:canc2}
$$
\|e^{iu_1 H_0}V(s_1) U(u_2)\|_{1\to \infty} \lesssim (|u_1|+
|u_2|)^{-\frac n2} e^{Mu_2}   \lesssim \langle u_1 \rangle^{-\frac
n2} \langle u_2 \rangle^{-\frac n2}e^{Mu_2}
$$
The case where $m>2$ follows exactly as the first inequality using
Lemma~\ref{lemma:canc1}.
\end{proof}

\section{Proof of the decay estimates}

\label{sec:decay} Theorem~\ref{thm:main} will be proved in this
section by a bootstrap argument. By Corollary~\ref{cor:shorttime},
we can assume that $t$ is large enough in Theorem~\ref{thm:main}.
More precisely, $t$ will be bigger than any constant appearing in
our estimate, except the bootstrap constant $C_0$ in
\eqref{eq:boot1}. By assumption, $ H_1, H_2$ can only admit
finitely many negative eigenvalues. Let $ \alpha > 0 $ satisfy:
$-\alpha$ is bigger than any eigenvalue of $ H_1, H_2$. For
technical reasons, we will assume that the initial data $\psi$
belong to $L^1 \cap L^2$ and employ the following bootstrap
argument:

Specifically, we will show that
\begin{equation}
\| U(t) \psi_0\|_{L^\infty }\leq C_0  |t|^{-\frac n2}
(\|\psi_0\|_{L^1} + e^{-\frac {\alpha T}2} \|\psi_0\|_{L^2})
\text{\ \ for all\ \ }0\le t\le T, \label{eq:boot1}
\end{equation} implies that
\begin{equation}
\|U(t) \psi_0\|_{L^\infty} \leq \frac{C_0}{2} |t|^{-\frac n2}
(\|\psi_0\|_{L^1} + e^{-\frac {\alpha T}2} \|\psi_0\|_{L^2})
\text{\ \ for all\ \ }0\le t\le T, \label{eq:boot2}
\end{equation}
provided that $\frac {C_0}2$ remains larger than some constant
that does not depend on $T$.  The logic here is that for arbitrary
but fixed $T$, the assumption~\eqref{eq:boot1} can be made to hold
for some $C_0$ depending on~$T$, because of
Corollary~\ref{cor:shorttime}. Iterating the implication
\eqref{eq:boot1}~$\Longrightarrow$~\eqref{eq:boot2} then yields a
constant that does not depend on~$T$. So we can let $T \to
+\infty$ to eliminate $\|\psi_0\|_{L^2}$ on the right-hand side.
Since $L^1 \cap L^2$ is dense in $L^1$ and $U(t)$ is a linear
operator, we get the dispersive estimate \eqref{eq:main} for any
initial data $\psi_0 \in L^1$. To simplify the notation, we write
$\|\psi_0\|_{L^1} + e^{-\frac {\alpha T}2} \|\psi_0\|_{L^2} $ as
$\trip\psi_0 \trip^{(T)}$ or $ \trip\psi_0 \trip$.

We proceed by expanding $U(t)$ via Duhamel's formula with respect
to the free evolution $H_0$:

\begin{align} \label{eq:DU1'} & U(t) \phi_0 = e^{-it H_0} \phi_0 - i
\int^t_0 e^{-i(t-s) H_0}V(s) U(s) \psi_0 \, ds \\ & = e^{-it H_0}
\psi_0 - i \int^t_0 e^{-i(t-s) H_0}V(s) e^{-is H_0} \psi_0 \, ds
\nn
\\ & \mbox{\hspace{.3in}}- \int^t_0 \int^s_0 e^{-i(t-s) H_0}V(s)
e^{-i(s-\tau) H_0}V(\tau)U(\tau) \psi_0 \, d\tau ds
\label{eq:DU2'}
\end{align}

Note that $ \|e^{-it H_0} \psi_0 \|_{\infty} \lesssim |t|^{-\frac
n2} \| \psi_0 \|_1 $. For the second term in~\eqref{eq:DU2'}, we
divide the integration interval $(0,t)$ into three pieces and
handle them by means of the cancellation lemma. Firstly,

$$ \| \int^1_0 e^{-i(t-s) H_0}V(s) e^{-is H_0} \psi_0 \, ds \|_{\infty}
\les |t|^{-\frac n2} \sup_s \|e^{is H_0}V(s) e^{-is H_0} \|_{1\to
1} \| \psi_0 \|_1 \les |t|^{-\frac n2} \| \psi_0 \|_1.
$$
Similarly, we have
$$ \| \int^t_{t-1} e^{-i(t-s) H_0}V(s) e^{-is H_0} \psi_0 \, ds \|_{\infty}
\les |t|^{-\frac n2} \sup_s \|e^{is H_0}V(s) e^{-is H_0} \|_{1\to
1} \| \psi_0 \|_1 \les |t|^{-\frac n2} \| \psi_0 \|_1.
$$
The third piece is

$$ \| \int^{t-1}_1 e^{-i(t-s) H_0}V(s) e^{-is H_0} \psi_0 \, ds \|_{\infty}
\les \int^{t-1}_1 |t-s|^{-\frac n2} \sup_s \|V(s)\|_1 |s|^{-\frac
n2} \, ds \| \psi_0\|_1  \les |t|^{-\frac n2} \| \psi_0 \|_1,
$$
where we observed that
\begin{equation}
\int^{t-1}_1 |t-s|^{-\frac n2}|s|^{-\frac n2} ds \les t^{-\frac
n2} \text{\ \ given\ \ } n \geq 3. \label{eq:integral}
\end{equation}

The third term in \eqref{eq:DU2'} is

\be \label{eq:hard1} \int^t_0 ds \int^s_0 d\tau \, e^{-i(t-s)
H_0}V(s) e^{-i(s-\tau) H_0}V(\tau) U(\tau) \psi_0.    \ee

We will decompose the domain of integration $\int^t_0 ds \int^s_0
d\tau$ into several pieces and treat each piece separately. We fix
$A>0$ as a large constant and $\epsilon>0$ as a small constant.
Write $\min \{s, A\}= s\wedge A $. Then Lemma~\ref{lemma:canc3}
and \eqref{eq:integral} implies that

\begin{align*} & \| \int^t_0 ds \int^{s \wedge A}_0 d\tau \, e^{-i(t-s) H_0}V(s)
e^{-i(s-\tau) H_0}V(\tau) U(\tau) \, d\tau ds \|_{1\to \infty}\\
&\les \int^t_0 ds \int^{s \wedge A}_0 d\tau \, \langle
t-s\rangle^{-\frac n2}\langle s-\tau \rangle^{-\frac n2} \langle
\tau \rangle^{-\frac n2} e^{AM}    \\ &\les t^{-\frac n2}.
\end{align*}

By $\|\cdot \|_{1\to \infty}$, we mean the operator norm from
$L^1$ to $L^\infty$. However when we apply the bootstrap
assumption, $\|\psi_0\|_{L^1}$ has to be modified to
$\|\psi_0\|_{L^1} + e^{-\frac {\alpha T}2}
\|\psi_0\|_{L^2}:=\trip\psi_0\trip$.

An application of Lemma~\ref{lemma:canc1} and the bootstrap
assumption show that
\begin{align*} & \| \int^t_{t-\epsilon} ds \int^s_{t -\epsilon} d\tau \, e^{-i(t-s)
H_0}V(s) e^{-i(s-\tau) H_0}V(\tau) U(\tau) \, d\tau ds \psi_0 \|_\infty \\
& \les  \int^t_{t-\epsilon} ds \int^s_{t -\epsilon} d\tau \,
\|e^{-i(t-s) H_0}V(s) e^{-i(s-\tau) H_0} \|_{1\to \infty}
\|V(\tau) U(\tau) \psi_0 \|_1  \, d\tau ds  \\
& \les \int^t_{t-\epsilon}d\tau \int^t_{\tau}ds |t-\tau|^{-\frac
n2} \, \max_{\tau \in (t-\epsilon,t)} \|U(\tau)\psi_0 \|_\infty .
\end{align*}

If $n=3$, then the above is dominated by

$$ \les \int^t_{t-\epsilon} d\tau |t-\tau|^{-\frac 12} C_0 t^{-\frac
n2} \trip\psi_0 \trip \les \sqrt{\epsilon} C_0 t^{-\frac
n2}\trip\psi_0 \trip.$$

Taking $\epsilon$ small enough, the above term can be dominated by
$\frac 1{100}C_0 t^{-\frac n2}\trip\psi_0 \trip $. When $n>3$, we
need to expand $U(t)$ further to remove the singularity of
$|t-\tau|^{-\frac n2}$ at $\tau=t$, (see~\cite{JSS} Section 2 for
details). The following is another piece of \eqref{eq:hard1}:

\begin{align*} & \| \int^{t-A}_A \int^s_A e^{-i(t-s) H_0}V(s) e^{-i(s-\tau)
H_0}V(\tau) U(\tau) \psi_0 \, d\tau ds \|_{\infty} \\ & \les
\int^{t-A}_A ds  \int^s_A d\tau \langle t-s\rangle^{-\frac n2}
\langle s-\tau \rangle^{-\frac n2} \| V(\tau) U(\tau) \psi_0 \|_1
\\ & \les \int^{t-A}_A ds  \int^s_A d\tau \langle
t-s\rangle^{-\frac n2} \langle s-\tau \rangle^{-\frac n2} C_0 \tau
^{-\frac n2} \trip\psi_0 \trip  \\ & \les  C_0 \trip \psi_0 \trip
\int^{t-A}_A ds \langle t-s\rangle ^{-\frac n2} \langle s
\rangle^{-\frac n2}  \\ & \les C_0 \trip \psi_0 \trip  t^{-\frac
n2} \kappa_A \leq \frac 1{100} C_0 \trip \psi_0 \trip  t^{-\frac
n2}.
\end{align*}

\noindent where $\kappa_A <\int^{+\infty}_A ds \langle s\rangle
^{-\frac n2} \to 0$ as $A \to \infty$. Lemma~\ref{lemma:canc3} and
the bootstrap assumption are applied in turn in the above. The
last line of above inequality holds provided that $A$ is large
enough. By Corollary~\ref{cor:shorttime}, we can assume $t >> A$.
Similarly, the following piece in \eqref{eq:hard1} also requires
that $A$ is large:

 \begin{align*} & \| \int^t_{t-A}
\int^{s-A}_A e^{-i(t-s) H_0}V(s) e^{-i(s-\tau) H_0}V(\tau) U(\tau)
\psi_0 \, d\tau ds \|_{\infty} \\
& \les \int^t_{t-A} ds  \int^{s-A}_A d\tau \langle t-s\rangle
^{-\frac n2} \langle s-\tau \rangle^{-\frac n2} \|U(\tau) \psi_0
\|_{\infty} \leq \frac 1{100} C_0 \trip \psi_0 \trip  t^{-\frac
n2}
\end{align*}

So By  what remains  in \eqref{eq:hard1} is

\be  \label{eq:hard2} \sum_{j=1}^m \int^t_{t-A} ds \int^{s \wedge
(t-\epsilon)}_{s-A} d\tau \, e^{-i(t-s) H_0}V(s) e^{-i(s-\tau)
H_0}V_j(\cdot-\tau \vec v_j ) U(\tau) \psi_0 \ee

For the term containing $V_j$ in \eqref{eq:hard2}, $U(\tau)$ will
be expanded with respect to $H_j$ by Duhamel's formula. Abusing
notation, we will write $V_1(\cdot - \tau \vec v_1)$ as
$V_1(\tau)$. In the following, we only deal with the term
containing $V_1$ which will be decomposed into two parts by
$U(\tau)=P_b(H_1,\tau)U(\tau)+P_c(H_1,\tau) U(\tau)$.

\subsection{Bound States}

\begin{prop}
\label{prop:bdstates} Let $\psi (t, x) = (U(t) \psi_0)(x)$ be a
solution of~\eqref{eq:2pot} which is asymptotically orthogonal to
the bound states of $H_j$, $j=1,2$ in the sense of
Definition~\ref{def:asymp}. Provided the bootstrap assumption
\eqref{eq:boot1}, we have for any $ t \in (0,T)$
\begin{equation}
\| P_b (H_1,t) U(t)\psi_0 \|_{\infty} \lesssim C_0 e^{-\frac
{\alpha t}4} t^{-\frac n2}( \|\psi_0\|_{L^1} + e^{-\frac {\alpha
T}2} \|\psi_0\|_{L^2}), \label{eq:bdstat}
\end{equation}

\noindent where $C_0$ is the constant in the bootstrap assumption.
\end{prop}

\begin{proof}

Let $\tilde U(t):= \calge1(t) U(t)$ and $\phi(t) = \tilde
U(t)\psi_0$. Then $\phi(t)$ solves \bea & \frac{1}{i} \partial_t
\phi - \Lapl \phi + V(\cdot+t\vec{v_1})
\phi = 0, \label{eq:dual} \\
& \phi|_{t=0}(x) = (\calge1(0) \psi_0)(x), \nn \eea

Then  $\|P_b(H_1,t) U(t)\psi_0\|_{\infty} = \|P_b (H_1 ) \tilde
U(t)\phi_0\|_{\infty} $ so without loss of generality, we can
assume that $ \vec{v_1}$ is the zero vector. Suppose that the
bound states of $H_1$ are $u_1, u_2,\ldots, u_l$ and we decompose
\begin{equation}
U(t) \psi_0 = \sum^l_{i=1} a_i (t) u_i + \psi_1 (t,x)
\label{eq:decomp}
\end{equation}
with respect to $H_1$ so that $P_c(H_1) \psi_1 = \psi_1$ and
$P_b(H_1)\psi_1 = 0.$ By the asymptotic orthogonality assumption,
$$
 \sum^l_{i=1} |a_i(t)|^2 \to 0 \text{ as } t \to \infty.
$$
Substituting \eqref{eq:decomp} into \eqref{eq:2pot} yields
\begin{align}
&\frac{1}{i}\partial_t \psi_1 - \Lapl \psi_1 + V_1 \psi_1 +
V_2(\cdot-t\vec{e_1}) \psi_1 +\nn \\
&+   \sum^l_{j=1} \left[\frac{1}{i} \dot a_j(t) u_j - \Lapl u_j
a_j (t) + V_1\, u_j a_j(t) +  V_2(\cdot - t\vec{e_1}) u_j
a_j(t)\right]=0. \label{eq:ai}
\end{align}

Since $P_c(H_1) \psi_1 = \psi_1$, we have
$$
(-\Lapl + V_1)\psi_1 = H_1 \psi_1 = P_c(H_1) H_1 \psi_1,\qquad
\partial_t \psi_1 = P_c(H_1) \partial_t
\psi_1.
$$
In particular,
$$P_b(H_1) \left(\frac{1}{i}\partial_t \psi_1 -
\Lapl\psi_1 + V_1 \psi_1\right) =0.
$$
Thus taking an inner product of the equation~\eqref{eq:ai} with
$u_\kappa$ and using the fact that $\langle u_\kappa , u_j\rangle
= \delta_{j\kappa}$ as well as the identity
$$
-\Lapl u_j + V_1 u_j = \lambda_j u_j,
$$
we obtain the ODE
$$
\frac{1}{i} \dot a_\kappa(t) + \lambda_\kappa a_\kappa(t) +
\langle V_2(\cdot - t \vec{e_1}) \psi_1, u_\kappa \rangle +
\sum_{j=1}^m a_j(t) \la V_2(\cdot-t\vec{e_1})u_j,u_\kappa \ra = 0
\label{eq:ODE}
$$
for each $a_\kappa$ with the condition that
$$
a_\kappa(t) \to 0 \text{ as } t \to + \infty.
$$
Recall that $u_\kappa$ is an eigenfunction of $H_1=-\Lapl + V_1$
with eigenvalue $\lambda_\kappa < 0$. It is well-known (see
e.g.~Agmon~\cite{Ag}) that such eigenfunctions are exponentially
localized, i.e.,
\begin{equation}
\label{eq:local} \int_{\R^n} e^{2\alpha |x|}\, |u_\kappa(x) |^2\,
dx  \le C= C(V_1,n) <\infty \text{ for some positive } \alpha.
\end{equation}
Therefore, the assumption that $V_2$ has compact support implies
\begin{equation}
\|V_2(\cdot - t \vec{e_1}) u_\kappa \|_2 \lesssim e^{-\alpha t}
\text{\ \ for all\ \ }t\ge 0 \label{eq:V2uk}.
\end{equation}
The implicit constant in~\eqref{eq:V2uk} depends on the size of
the support of $V_2$ and $\| V_2\|_{L^\infty}$.

By the bootstrap assumption, $ f_\kappa(t) : = \la V_2(\cdot -
t\vec{e_1})\psi_1, u_\kappa \ra $ satisfies

\begin{align}
 |f_\kappa(t)| & \lesssim \|\psi_1 \|_{\infty} \|V_2(\cdot -
t\vec{e_1}) u_\kappa\|_1 \lesssim  e^{-\alpha t}\|\psi_1
\|_{\infty} \nn  \\
& \lesssim e^{-\alpha t}\|(Id-P_b(H_1)) U(t)\psi_0
\|_{\infty}  \nn \\
& \lesssim e^{-\alpha t} t^{-\frac n2}C_0 (\|\psi_0\|_{L^1} +
e^{-\frac {\alpha T}2} \|\psi_0\|_{L^2}) + e^{-\alpha t}
\sum^l_{i=1} |a_i (t)| \, \|u_i\|_{\infty}, \label{eq:fkappa}
\end{align}
\noindent where $t \in (0,T)$. Notice that \eqref{eq:fkappa} fails
for $t > T$ because the bootstrap assumption only applies to
$0<t<T $. Instead, we have the following for $t > T$:

\be |f_\kappa(t)| < \|V_2(\cdot - t\vec{e_1})u_\kappa\|_2
\|\psi_1\|_2 \lesssim e^{-\alpha t} \|\psi_0\|_2.
\label{eq:fkappaT}\ee

In view of~\eqref{eq:ODE},  $a_\kappa$ solves the equation \bea &
\frac{1}{i} \dot a_\kappa (t) + \lambda_\kappa a_\kappa(t) +
\sum_{j=1}^m a_j(t) C_{j\kappa}(t) + f_\kappa(t) =0 \label{eq:ODE2} \\
& a_\kappa(\infty) = 0,\nn \eea where $C_{j\kappa}(t)=C_{\kappa
j}(t)=\la V_2(\cdot-t\vec{e_1})u_j,u_\kappa \ra$.
By~\eqref{eq:V2uk}, $\max_{j,\kappa} |C_{j\kappa}(t)| \lesssim
e^{-\alpha t}.$ Solving \eqref{eq:ODE2} explicitly, we obtain
$$
\vec a (t) = ie^{-i \int_0^t B(s)\,ds}\int^\infty_t e^{i\int_0^s
B(\tau)\,d\tau} \vec f(s) \,ds,
$$
where $B_{j\kappa}(t)=\lambda_j\delta_{j\kappa}+C_{j\kappa}(t)$.

By \eqref{eq:fkappa}, \eqref{eq:fkappaT} and the unitarity of
$e^{i\int_0^s B(\tau)\, d\tau}$, we conclude that
\begin{align*}
& |\vec a(t) | \le \int_t^T + \int_T^\infty |\vec f(s) |\, ds \\
\lesssim \int_t^T e^{-\alpha s} s^{-\frac n2} & C_0 ds \trip
\psi_0\trip  + \int_t^T e^{-\alpha s} \sum^l_{j=1} |a_j(s)|
\|u_i\|_{\infty} ds + \int_T^\infty e^{-\alpha s} ds
\|\psi_0\|_{L^2}
\end{align*}

Choose a large constant $t_0>0$ such that for all $t_1>t_0$, the
following holds:

\be  \int_{t_1}^T e^{-\alpha s} \sum^l_{j=1} |a_j(s)|
\|u_i\|_{\infty} ds \leq \frac12 \sup_{t_1<t<T}|\vec a(t)| ,\ee

\noindent then $$ \sup_{t_1<t<T}|\vec a(t)| \lesssim e^{-\alpha
t_1} t_1^{-\frac n2} C_0 \trip \psi_0\trip  + e^{-\alpha T}
\|\psi_0\|_{L^2} \lesssim  e^{-\frac {\alpha t_1}4} t_1^{-\frac
n2} C_0 \trip \psi_0\trip  $$

\end{proof}

\begin{remark} \label{rem:bdsates}
In the above proof, if we change \eqref{eq:fkappa} into the
following:

\begin{align*}
 & |f_\kappa(t)| \lesssim \|\psi_1(t) \|_p \|V_2(\cdot -
t \vec{e_1}) u_\kappa\|_{p'} \lesssim  e^{-\alpha t}\|\psi_1(t)
\|_p
\\ & \lesssim e^{-\alpha t}\|\ (Id-P_b(H_1)) U(t)\psi_0
\|_p  \\
& \lesssim e^{-\alpha t} t^{-\gamma}C_0 (\| \psi_0 \|_{p'} +
e^{-\frac {\alpha T}2} \|\psi_0\|_{L^2} )  + e^{-\alpha t}
\sum^l_{i=1} |a_i (t)| \, \|u_i\|_p \label{eq:fkappa2}
\end{align*}

\noindent where $\gamma=n(\frac 12-\frac 1p) >1$, and follow the
same arguments, we see that for large $t$,

$$\| P_b(H_1,t)U(t)
\psi_0\|_p \lesssim t^{-\gamma}C_0 e^{-\frac {\alpha T}4}(
\|\psi_0 \|_{p'}+ e^{-\frac {\alpha T}2} \|\psi_0\|_{L^2} ).$$

If the potential $V_1$ is smooth enough, it is known (see e.g.
\cite{Ag}) that the bound state $u_j$ of $H_1$ is differentiable.
Moreover, its derivatives decay exponentially at infinity. Thus,
\begin{equation} \label{eq:bdstat2}
 \| \partial P_b(H_1,t)U(t)
\psi_0\|_p \leq \sum^l_{i=1} |a_i (t)| \|\partial u_i\|_p \lesssim
t^{-\gamma}C_0 e^{-\frac {\alpha T}4}( \|\psi_0 \|_{p'}+ e^{-\frac
{\alpha T}2} \|\psi_0\|_{L^2} ).
\end{equation}
In addition, the above claims hold with $H_1$ replaced by $H_j$,
$j=2,\cdots,m$. These results will be used to prove
Theorem~\ref{thm:main2} in Section 4. $\square$
\end{remark}

With Proposition~\ref{prop:bdstates}, the $P_b(H_1,\tau)U(\tau)$
part of \eqref{eq:hard2} can be estimated by the following:

\begin{align*} & \int^t_{t-A} ds \int^{s \wedge (t-\epsilon)}_{s-A} d\tau
\|e^{-i(t-s) H_0}V(s) e^{-i(s-\tau) H_0}V_1(\tau) P_b(H_1,\tau)
U(\tau) \psi_0 \|_{\infty} \\
\lesssim & \int^t_{t-A} ds \int^{s \wedge (t-\epsilon)}_{s-A}
d\tau \la t-s \ra^{-\frac n2} \la s-\tau \ra^{-\frac n2} \|
V_1(\tau) P_b(H_1,\tau) U(\tau) \psi_0 \|_1 \\
\lesssim & A^2  \sup_{\tau \in (t-2A,t)} \|
P_b(H_1,\tau)U(\tau)\psi_0\|_{\infty} \\
& < \frac {C_0}{100} t^{-\frac n2} (\|\psi_0\|_{L^1} + e^{-\frac
{\alpha T}2} \|\psi_0\|_{L^2}). \end{align*}

For the $P_c(H_1,\tau)U(\tau)$ part of \eqref{eq:hard2}, we need
to apply Duhamel's formula again and expand \eqref{eq:hard2}
further with respect to $H_1$. We assume that $\vec {v_1}=0$ and
$m=2$ to simplify our notation. Specifically, we plug the
following
$$P_c(H_1,\tau)U(\tau)= P_c(H_1)U(\tau)= P_c(H_1)e^{-i\tau H_1}-i
P_c(H_1) \int_0^{\tau} e^{-i(\tau-r) H_1}V_2(r)U(r)\, dr$$

\noindent into \eqref{eq:hard2}. For the term containing
$P_c(H_1)e^{-i\tau H_1}$, we apply the dispersive decay for
$P_c(H_1)e^{-i\tau H_1}$:

\bea \int^t_{t-A} ds \int^{s \wedge (t-\epsilon)}_{s-A} d\tau
\|e^{-i(t-s) H_0}V(s) e^{-i(s-\tau) H_0}V_1(\tau) P_c(H_1)e^{-i\tau H_1}\psi_0 \|_\infty \nn \\
\lesssim \int^t_{t-A} ds \int^{s \wedge (t-\epsilon)}_{s-A} d\tau
\langle t-s\rangle^{-n/2} \langle s-\tau \rangle^{-n/2}\langle
\tau \rangle^{-n/2} \|\psi_0\|_1 \lesssim t^{-n/2}\|\psi_0\|_1.
\nn \eea

The second term is \be  \label{eq:hard2Pc}
 \int^t_{t-A} ds \int^{s \wedge (t-\epsilon)}_{s-A} d\tau
\int_0^{\tau}dr \, e^{-i(t-s) H_0}V(s) e^{-i(s-\tau) H_0}V_1(\tau)
P_c(H_1) e^{-i(\tau-r) H_1}V_2(r)U(r) \psi_0 .\ee

Now take a small constant $\delta >0$ and a large constant $A_1>0$
to be specified later. We decompose the integral $\int_0^{\tau}dr
$ in \eqref{eq:hard2Pc} as following:

\be  \label{eq:decomp2} \int_0^{\tau}dr = \int_0^{\delta}dr +\int_
{\delta}^{A_1} dr+\int_{A_1}^{\tau-A_1}dr +
\int_{\tau-A_1}^{\tau-\delta}dr + \int_{\tau-\delta}^{\tau}dr .\ee

To simplify the notation, we will write $A_1$ as $A$. Our goal is
to estimate each term in \eqref{eq:decomp2}. The second term of
\eqref{eq:decomp2} is estimated as follows:

\begin{align*}
& \int^t_{t-A} ds \int^{s \wedge (t-\epsilon)}_{s-A} d\tau
\int_{\delta}^A dr \|e^{-i(t-s) H_0}V(s) e^{-i(s-\tau)
H_0}V_1(\tau)
P_c(H_1) e^{-i(\tau-r) H_1}V_2(r)U(r) \psi_0 \|_{\infty} \\
& \lesssim \int^t_{t-A} ds \int^{s \wedge (t-\epsilon)}_{s-A}
d\tau \int_{\delta}^A dr \langle t-s\rangle^{-n/2} \langle s-\tau
\rangle^{-n/2}\langle \tau -r \rangle^{-n/2} \langle r
\rangle^{-n/2} e^{rM} \|\psi_0\|_1 \\
& \lesssim t^{-n/2} \|\psi_0\|_1.
\end{align*}
The implicit constant above depends on $A , \delta$.

The third term of \eqref{eq:decomp2} is estimated as follows:
\begin{align*}& \int^t_{t-A} ds \int^{s \wedge (t-\epsilon)}_{s-A} d\tau \int_A^{\tau-A} dr
\|e^{-i(t-s) H_0}V(s) e^{-i(s-\tau) H_0}V_1(\tau) P_c(H_1)
e^{-i(\tau-r) H_1}V_2(r)U(r) \psi_0\|_{\infty} \\
& \lesssim \int^t_{t-A} ds \int^{s \wedge (t-\epsilon)}_{s-A}
d\tau \int_A^{\tau-A} dr \langle t-s\rangle^{-n/2} \langle s-\tau
\rangle^{-n/2}\langle \tau -r \rangle^{-n/2}
\|U(r)\psi_0\|_{\infty}\\
& \lesssim \int^t_{t-A} ds \int^{s \wedge (t-\epsilon)}_{s-A}
d\tau \int_A^{\tau-A} dr \langle t-s\rangle^{-n/2} \langle s-\tau
\rangle^{-n/2}\langle \tau -r \rangle^{-n/2} \langle
r\rangle^{-n/2} C_0
\trip \psi_0\trip \\
 & \lesssim t^{-n/2} C_0 \kappa_A \trip \psi_0\trip
\leq  \frac 1{100}C_0 t^{-n/2} \trip \psi_0\trip ,
\end{align*}
where $\kappa_A \to 0$ as $A \to \infty$. So the above inequality
holds for large enough $A$.

For the fourth term in \eqref{eq:decomp2}, we have:
\begin{align*}&  \int^t_{t-A} ds \int^{s \wedge (t-\epsilon)}_{s-A} d\tau \int_{\tau- \delta}^{\tau} dr
\|e^{-i(t-s) H_0}V(s) e^{-i(s-\tau) H_0}V_1(\tau)
e^{-i(\tau-r) H_1}P_c(H_1) V_2(r)U(r) \psi_0\|_{\infty} \\
& \lesssim \int^t_{t-A} ds \int^{s \wedge (t-\epsilon)}_{s-A}
d\tau \int_{\tau-A}^\tau dr \langle t-s\rangle^{-n/2} \langle
s-\tau \rangle^{-n/2} \|V_1(\tau)P_c(H_1)e^{-i(\tau-r) H_1} V_2(r)U(r)\psi_0\|_1\\
& \lesssim \int^t_{t-A} ds \int^{s \wedge (t-\epsilon)}_{s-A}
d\tau \int_{\tau-A}^\tau dr \langle t-s\rangle^{-n/2} \langle
s-\tau \rangle^{-n/2} \|U(r)\psi_0\|_{\infty}\\
& \lesssim \int^t_{t-A} ds \int^{s \wedge (t-\epsilon)}_{s-A}
d\tau \int_{\tau- \delta}^{\tau} dr \langle t-s\rangle^{-n/2}
\langle s-\tau \rangle^{-n/2} \langle r \rangle^{-n/2} C_0  \trip \psi_0\trip \\
& \lesssim t^{-n/2} C_0 \kappa_{\delta} \trip \psi_0\trip  \leq
\frac 1{100}C_0 t^{-n/2} \trip \psi_0\trip ,
\end{align*}
where $\kappa_{\delta} \to 0$ as $\delta \to 0$. So the above
inequality holds for $\delta$ small enough.

For the $\int_0^{\delta}\, dr$ part of \eqref{eq:decomp2}, we
expand
 $$e^{-i(\tau-r) H_1} = e^{-i(\tau-r) H_0}
-i\int_0^{\tau-r} e^{-i(\tau-r-\beta) H_1}V_1 e^{-i\beta H_0} \,
d\beta . $$

Here we put $H_0$ after $H_1$ in the integral because we want
$H_0$ to appear immediately before $U(r)$ and apply
Lemma~\ref{lemma:canc3}. Substitute this expansion into the
$\int_0^{\delta}\, dr$ part of \eqref{eq:decomp2} and we get two
terms. The first one is

\be \int^t_{t-A} ds \int^{s \wedge (t-\epsilon)}_{s-A} d\tau
\int_0^{\delta} dr e^{-i(t-s) H_0}V(s) e^{-i(s-\tau) H_0}V_1(\tau)
P_c(H_1) e^{-i(\tau-r) H_0}V_2(r)U(r) \psi_0  \label{eq:hard31}\ee

Notice that $P_c(H_1)=Id-P_b(H_1)$, and because $\|P_b(H_1)\|_{p
\to p}$ is bounded, $\|P_c(H_1)\|_{p \to p}$ is bounded as well.
Therefore the $L^{\infty}$ norm of \eqref{eq:hard31} is estimated
as follows:

$$ \lesssim \int^t_{t-A} ds \int^{s \wedge (t-\epsilon)}_{s-A}
d\tau \int_0^{\delta} dr \langle t-s\rangle^{-n/2} \langle s-\tau
\rangle^{-n/2} \langle \tau \rangle^{-n/2} e^{Mr}\|\psi_0\|_1
\lesssim t^{-n/2} \|\psi_0\|_1.
$$

The second term  of the $\int_0^{\delta}\, dr$ part of
\eqref{eq:decomp2} after substitution is

\begin{align}& \int^t_{t-A} ds \int^{s \wedge (t-\epsilon)}_{s-A}
d\tau \int_0^{\delta} dr \, e^{-i(t-s) H_0}  V(s) e^{-i(s-\tau)
H_0}V_1(\tau)\cdot  \nn \\ & \mbox{\hspace{.5in}} P_c(H_1)
\int_0^{\tau-r} e^{-i(\tau-r-\beta) H_1}V_1 e^{-i\beta H_0} \,
d\beta \, V_2(r)U(r) \psi_0  \label{eq:hard32} \end{align}

Decompose  $\int_0^{\tau-r} d\beta $ so we can rewrite
\eqref{eq:hard32}$= J_1 + J_2+ J_3$, where $J_1,J_2 $ and $J_3$
correspond to $\int_ 0^{\delta} d\beta$, $\int_{\delta}^{\tau-r-1}
d\beta $ and $\int_ {\tau-r-1}^{\tau-r} d\beta $ respectively.

We proceed to estimate $J_1$ as follows:

\begin{align*} & \int_0^{\delta} dr \int_0^{\delta} d\beta \,
\|P_c(H_1) e^{-i(\tau-r-\beta) H_1}V_1 e^{-i\beta H_0}
V_2(r)U(r)\psi_0 \|_{\infty} \\
& \lesssim  \int_0^{\delta} dr \int_0^{\delta} d\beta \, \langle
\tau-r-\beta \rangle^{-n/2} \|e^{-i\beta H_0} V_2(r)U(r)\psi_0\|_{\infty} \\
& \lesssim \langle \tau \rangle^{-n/2} \int_0^{\delta} dr
\int_0^{\delta} d\beta (\beta+r)^{-n/2}e^{Mr} \|\psi_0\|_1.
\end{align*}

In the above expression, when $n=3$, $\int_0^{\delta} dr
\int_0^{\delta} d\beta (\beta+r)^{-n/2}e^{Mr}$ is integrable. When
$n>3$, we need to further expand $e^{-i(\tau-r-\beta) H_1}$ to
remove the singularity of $(\beta+r)^{-n/2}$ at $\beta+r=0$. In
either case, we can conclude that $\|J_1\|_\infty \lesssim
t^{-n/2} \|\psi_0\|_1 $.

 For $J_2$, our estimate is the following:

\begin{align*}& \int_0^{\delta} dr \int_{\delta}^{\tau-r-1} d\beta
\|P_c(H_1) e^{-i(\tau-r-\beta) H_1}V_1 e^{-i\beta H_0}
V_2(r)U(r)\psi_0 \|_{\infty} \\
& \lesssim \int_0^{\delta} dr \int_{\delta}^{\tau-r-1} d\beta
\langle \tau-r-\beta \rangle^{-n/2} \|V_1 e^{-i\beta H_0}
V_2(r)U(r)\psi_0\|_1 \\
& \lesssim \int_0^{\delta} dr \int_{\delta}^{\tau-r-1} d\beta
\langle \tau-r-\beta \rangle^{-n/2} \|e^{-i\beta H_0}
V_2(r)U(r)\psi_0\|_{\infty} \\
& \lesssim \int_0^{\delta} dr \int_{\delta}^{\tau-r-1} d\beta
\langle \tau-r-\beta \rangle^{-n/2} (\beta+r)^{-n/2}e^{Mr}\|\psi_0\|_1\\
& \lesssim \int_0^{\delta} dr \int_{\delta}^{\tau-r-1} d\beta
\langle \tau-r-\beta \rangle^{-n/2}\langle \beta +r \rangle^{-n/2} \|\psi_0\|_1\\
& \lesssim \tau^{-n/2} \| \psi_0\|_1 . \end{align*}

\noindent The implicit constant above depends on $\delta$ and is
independent of $t$ and $\psi_0$. Plugging the above estimate into
$J_1$, we derive that $\|J_2\|_\infty \lesssim t^{-n/2}
\|\psi_0\|_1 $.

To estimate $J_3$, we notice that

\begin{align*}& \|\int_ {\tau-r-1}^{\tau-r} d\beta V_1(\tau)
P_c(H_1)
e^{-i(\tau-r-\beta) H_1}V_1 e^{-i\beta H_0} V_2(r)U(r)\psi_0 \|_1 \\
& \leq \int_ {\tau-r-1}^{\tau-r} d\beta \|V_1(\tau)\|_2 \|P_c(H_1)
e^{-i(\tau-r-\beta) H_1} \|_{2\to 2}\|V_1 e^{-i\beta H_0}
V_2(r)U(r)\psi_0 \|_2 \\
& \lesssim \int_ {\tau-r-1}^{\tau-r} d \beta \|V_1\|_2
\|e^{-i\beta
H_0} V_2(r)U(r)\psi_0 \|_{\infty} \\
& \lesssim \int_ {\tau-r-1}^{\tau-r} d \beta |\beta|^{-\frac
n2}e^{Mr} \|\psi_0 \|_1.
\end{align*}

\noindent Observe that $r$ is small and $\beta \simeq \tau $.
Plugging the above estimate into $J_3$, we derive that
$\|J_3\|_\infty \lesssim t^{-\frac n2} \|\psi_0 \|_1$. Thus we
finished the estimate of the $\int_0^{\delta}dr$ part of
\eqref{eq:decomp2}.

\subsection{Low and high velocity estimates}
So far we have estimated four parts of \eqref{eq:decomp2}. This
subsection is devoted to deriving the estimate of the
$\int_{\tau-A}^{\tau- \delta}dr$ part of \eqref{eq:decomp2}, which
will be decomposed as follows:

\begin{align*} &  \int^t_{t-A} ds \int^{s \wedge
(t-\epsilon)}_{s-A} d\tau \int_{\tau-A}^{\tau-\delta}dr
 e^{-i(t-s)H_0}V(s) e^{-i(s-\tau) H_0}
V_1 P_c(H_1) e^{-i(\tau-r) H_1}V_2(r)U(r) \psi_0 \\
&=\int^t_{t-A} ds \int^{s \wedge (t-\epsilon)}_{s-A} d\tau
\int_{\tau-A}^{\tau-\delta}dr
 e^{-i(t-s)H_0}V(s) e^{-i(s-\tau) H_0} \cdot \\
& \mbox{\hspace{.5in}} \quad V_1 P_c(H_1) e^{-i(\tau-r) H_1}(F(|\vec p|\geq N)+F(|\vec p| \leq N))V_2(r)U(r) \psi_0 \\
&= J_{high} + J_{low}. \label{eq:hard4} \end{align*}

\noindent $F( |\vec p|\leq N)$ and $F(|\vec p|\geq N)$ denote
smooth projections onto the frequencies $|\vec p| \leq N$ and
$|\vec p| \geq N$, respectively. For the low velocity part
$J_{low}$, firstly, $(t-s)+(s-\tau) \geq \epsilon$ and
Lemma~\ref{lemma:canc3} imply

\be \label{eq:canc3'} \|e^{i(t-s) H_0}V(s) e^{i(s-\tau)
H_0}\|_{1\to \infty} \lesssim \langle t-s \rangle^{-\frac n2}
\langle s-\tau \rangle^{-\frac n2}. \ee

Secondly, we need the following proposition (see~\cite{RSS1} for
its proof):

\begin{prop} \label{prop:lowvel}
Let $\chi_{\tau}$ be a smooth cut of $B(0,\tau\delta)$, where
$\delta$ is a small constant depending only on $\vec{e_1}$ and
$B(0,\tau\delta)$ is a ball in $R^n$ centered at $0$ with radius
$\tau\delta$. Let $A, N$ be large positive constants and $A, N <<
r$ then
$$
\sup_{0< \tau-r \leq A} \| \chi_{\tau} e^{-i(\tau-r)H_1} P_c(H_1)
F(|\vec p|\leq N) V_2 (\cdot-r\vec{e_1}) \|_{L^2\to L^2} \leq
\frac{AN}{\delta t}.
$$
\end{prop}

The idea behind Proposition~\ref{prop:lowvel} can be explained as
the following:

The support of $V_2(\cdot - r \vec{e_1})$ is contained in
$B(r\vec{e_1},R)$. Here $R$ is the size of the support of $V_2$.
The operator $e^{-i(\tau-r)H_1} P_c (H_1) F( |\vec p| \leq N)$ can
``propagate'' $B(r\vec{e_1},R)$ into $B(0,\tau\delta)$ only if
$(\tau-r) N \ge \text{dist} (B(r\vec{e_1},R), B(0,\tau\delta))$
according to the classical picture. However if $|\tau -r|<A$,
$\tau \ll A, N$,  $(\tau-r) N \ll \text{dist} (B(r\vec{e_1},R),
B(0,\tau\delta))$ .

To apply this proposition to $J_{low}$, note that $\chi_{\tau} V_1
= V_1$. Let $\chi_2$ be a smooth cut of the support of $V_2$ and
$f$ be any function in $ L^\infty(R^n)$. Then it follows from
Proposition~\ref{prop:lowvel} that

\begin{align*}& \quad   \|V_1 P_c(H_1) e^{-i(\tau-r) H_1}F(|\vec p| \leq N)V_2(r)f\|_1\\
&=\|V_1 \chi_{\tau} P_c(H_1) e^{-i(\tau-r) H_1}F(|\vec p| \leq N)
V_2(r)\chi_2(\cdot- r \vec v_2)  f \|_1 \\
& \leq \|V_1\|_2 \| \chi_{\tau} P_c(H_1) e^{-i(\tau-r)
H_1}F(|\vec p| \leq N) V_2(r)\|_{2\to 2} \|\chi_2\|_2 \|f\|_\infty \\
& \lesssim \frac{ANM^2}{\delta t} \|f\|_\infty.
\end{align*}

Combining the above estimate with \eqref{eq:canc3'} and noting
$A,M,N \ll t$, we conclude
$$\|J_{low}\|_\infty \lesssim \int^t_{t-A} ds \int^{s \wedge (t-\epsilon)}_{s-A}
 d\tau \int_{\tau-A}^{\tau-\delta}dr
  \langle t-s\rangle^{-n/2} \langle s-\tau \rangle^{-n/2}
\frac{ANM^2}{\delta t} \|U(r) \psi_0\|_{\infty} \leq \frac
{C_0}{100} t^{-n/2} \trip \psi_0\trip .
$$

From the above estimate for $J_{low}$, it is worth remarking that
the purpose of the multiple expansions by Duhamel's formula is to
prepare a cushion (the potentials $ V_1$ and $ V_2$) to apply the
$L^2 \to L^2 $ estimate (Prop~\ref{prop:lowvel}) between $L^1 \to
L^\infty $ estimates.

For the high velocity part $J_{high}$, we shall further expand
$U(r)$ with respect to $H_0$, followed by a commutator argument.
By Duhamel's formula

$$U(r)= e^{-ir H_0}-i\int^r_0 e^{-i(r-\alpha)H_0}
V(\alpha)U(\alpha) \, d\alpha, $$
 we write $J_{high}= J_{high,1}-i J_{high,2}$, where

$$ J_{high,1}=\! \int^t_{t-A} \! ds \! \int^{s \wedge (t-\epsilon)}_{s-A}
\!\! d\tau  \! \int_{\tau-A}^{\tau-\delta} \! dr \,
 e^{-i(t-s)H_0}V(s) e^{-i(s-\tau) H_0}
V_1 P_c(H_1) e^{-i(\tau-r) H_1}F(|\vec p|\geq N)V_2(r)e^{-ir
H_0}\psi_0 ,$$

and
\begin{align*} J_{high,2}=\int^t_{t-A} ds \int^{s \wedge (t-\epsilon)}_{s-A} d\tau &
\int_{\tau-A}^{\tau-\delta}dr \, e^{-i(t-s)H_0}V(s) e^{-i(s-\tau)
H_0} V_1 P_c(H_1) e^{-i(\tau-r) H_1}F(|\vec p|\geq N) \cdot \\
& \cdot V_2(r) \int^r_0 e^{-i(r-\alpha)H_0}
V(\alpha)U(\alpha)\psi_0 \, d\alpha .
\end{align*}
The decay of $J_{high,1} $ will come easily from $e^{-irH_0}$.
Indeed, we apply Lemma~\ref{lemma:canc3} to $e^{-i(t-s)H_0}V(s)
e^{-i(s-\tau) H_0}$ as in \eqref{eq:canc3'} and notice that

\be \|P_c(H_1) e^{-i(\tau-r) H_1}F(|\vec p|\geq N)\|_{L^2 \to
L^2}\leq 1 . \label{eq:bound22}\ee
Then it is clear that $
\|J_{high,1}\|_{\infty}$ is dominated by
$$ \int^t_{t-A} ds \int^{s \wedge (t-\epsilon)}_{s-A} d\tau
\int_{\tau-A}^{\tau-\eta} dr \langle t-s\rangle^{-n/2} \langle
s-\tau \rangle^{-n/2}\langle r \rangle^{-n/2}\|\psi_0\|_1 \lesssim
t^{-n/2}\|\psi_0\|_1.
$$

$J_{high,2}$ will be decomposed into three parts $J_{high,2}^1,
J_{high,2}^2$ and $J_{high,2}^3$, corresponding to $\int^B_0
d\alpha$, $\int^{r-B}_B d\alpha$ and $\int^r_{r-B} d\alpha$
respectively, where $B>0$ is a large constant to be specified.

For $J_{high,2}^1$, the decay comes from $e^{-i(r-\alpha)H_0} $.
Indeed, it follows from Lemma~\ref{lemma:canc2} and $0< \alpha < B
$ that

$$ \|e^{-i(r-\alpha)H_0}V(\alpha)U(\alpha)\|_{1\to\infty}
 \lesssim r^{-n/2}e^{M\alpha} \lesssim \langle r \rangle^{-n/2}
$$

Hence, it follows from \eqref{eq:canc3'}, \eqref{eq:bound22} and
the above inequality that $ \|J_{high,2}^1\|_{\infty}$ is
dominated by
$$
\int^t_{t-A} ds \int^{s \wedge
(t-\epsilon)}_{s-A} d\tau \int_{\tau-A}^{\tau-\delta}dr
\langle t-s\rangle^{-n/2} \langle s-\tau \rangle^{-n/2} \langle r
\rangle^{-n/2}\|\psi_0\|_1 \lesssim \langle t \rangle^{-n/2}
\|\psi_0\|_1 .$$

$J_{high,2}^2$ will be estimated by an application of the
bootstrap assumption and the smallness comes from choosing $B$
sufficiently large. Indeed, it follows from \eqref{eq:canc3'},
\eqref{eq:bound22}, Lemma~\ref{lemma:canc3} and the bootstrap
assumption that

\begin{align*} & \|J_{high,2}^2\|_{\infty} \lesssim \int^t_{t-A} ds \int^{s
\wedge (t-\epsilon)}_{s-A} d\tau \int_{\tau-A}^{\tau-\delta}dr
\langle t-s\rangle^{-n/2} \langle s-\tau \rangle^{-n/2} \cdot \\ &
\mbox{\hspace{.7in}}\qquad \cdot \int^{r-B}_B \langle r-\alpha
\rangle
^{-n/2} \langle \alpha \rangle^{-n/2} d\alpha \, C_0 \trip \psi_0\trip  \\
& \lesssim \int^t_{t-A} ds \int^{s \wedge (t-\epsilon)}_{s-A}
d\tau \int_{\tau-A}^{\tau-\delta}dr \langle t-s\rangle^{-n/2}
\langle s-\tau \rangle^{-n/2} \langle r \rangle^{-n/2} \kappa_B
C_0
\trip \psi_0\trip  \\
& \leq \frac 1{100}C_0 t^{-n/2} \trip \psi_0\trip .
\end{align*}

\noindent In the above inequality, $B$ is chosen sufficiently
large, because $\kappa_B = \int_B^\infty \langle \alpha
\rangle^{-n/2} d\alpha \to 0$ when $B \to \infty$.

The decay of $J_{high,2}^3$ can only come from $U(\alpha)$. As
usual we need to generate the smallness $\frac 1{100}$ for the
bootstrap assumption. Here the smallness $\frac 1{100}$ comes from
the high velocity and a commutator argument. Write $F(|\vec p|\geq
N)V_2(r)= [F(|\vec p|\geq N),V_2(r)]+V_2(r) F(|\vec p|\geq N)$ and
correspondingly, we decompose $J_{high,2}^3 =J_{high,2}^{3,1}+
J_{high,2}^{3,2}$. That is to say $J_{high,2}^{3,1}$ and
$J_{high,2}^{3,2}$ are just $J_{high,2}^3 $ with $F(|\vec p|\geq
N)V_2(r)$ replaced by $[F(|\vec p|\geq N),V_2(r)]$ and $V_2(r)
F(|\vec p|\geq N)$.

Specifically, the smallness $\frac 1{100}$ for $J_{high,2}^{3,1}$
comes from the following standard fact, namely
\begin{equation}
\|[F(|\vec p|\leq N),V_2]\|_{L^2 \to L^2}\lesssim N^{-1}\,\|\nabla
V_2 \|_\infty .\label{eq:3comm}
\end{equation}

To see this, write $F(|\vec p|\leq N) f = [\hat{\eta}(\xi/N)
\hat{f}(\xi)]^{\vee}$ with some smooth bump function~$\eta$. Hence
the kernel $K$ of $[F(|\vec p|\leq N),V_2]$ is
\[ K(x,y) = N^n\eta(N(x-y))(V_2(y)-V_2(x)),\]
and \eqref{eq:3comm} follows from Schur's test and
$\sup_{x}\|K(x,\cdot)\|_{L^1} = \sup_{y}\|K(\cdot, y)\|_{L^1}
\lesssim N^{-1}\,\|\nabla V_2 \|_\infty $.

It follows from \eqref{eq:canc3'}, $\|P_c(H_1) e^{-i(\tau-r)
H_1}\|_{2\to 2} \leq 1 $, \eqref{eq:3comm} and the bootstrap
assumption that

\begin{align*}& \,\quad \|J_{high,2}^{3,1}\|_{\infty} \\
& \lesssim \int^t_{t-A} \! ds \! \int^{s \wedge
(t-\epsilon)}_{s-A} \! \! d\tau \int_{\tau-A}^{\tau-\delta}dr (\la
t-s\ra \la s-\tau \ra)^{-n/2} \frac {\|V_1\|_2 \|\nabla
V_2\|_\infty} N \int ^r_{r-B} \| e^{-i(r-\alpha)H_0}
V(\alpha)U(\alpha)
 \psi_0\|_2d\alpha \\
 & \lesssim \frac 1N \sup_{t-3A-B< \alpha< t} \|U(t)\psi_0\|_\infty
 \lesssim \frac { C_0}N t^{-n/2} \trip \psi_0\trip  \leq \frac {C_0}{100}  t^{-n/2}
 \trip \psi_0\trip ,
\end{align*}

\noindent where $\frac 1N$ is chosen sufficiently small to
dominate the implicit constant in $``\lesssim ''$ which only
depends on $n,V, \vec v_2$ and $\epsilon,\delta,A,B$.

The smallness for $J_{high,2}^{3,2}$ comes from the following
version of Kato's $\frac 12 -$smoothing estimate:

\be \|\int^{\alpha + B}_{\alpha} \chi_2(\cdot- r\vec v_2) F(|\vec
p|\geq N)
 e^{-i(r-\alpha)H_0}\, dr \|_{2 \to 2}
 \lesssim \frac {BR}{\sqrt{N}} \, ,  \label{eq:Kato}
 \ee
where $\chi_2(\cdot)$ is a smooth cut around the support of $V_2$
and $R$ is radius of the support of $\chi_2$. The implicit
constant only depends on $n, V_2$. We refer to \cite{RSS1} Section
$3.5$ for its proof and further references.

Now observe that the region of integration
$\int_{\tau-A}^{\tau-\delta}dr \int_r^{r-B}d\alpha$ is contained
in that of $\int_{\tau-A-B}^{\tau-\delta}d\alpha
\int_{\alpha}^{\alpha+B}dr$ and $\|P_c(H_1) e^{-i(\tau-r)
H_1}\|_{2\to 2} \leq 1 $. It follows from \eqref{eq:canc3'},
\eqref{eq:Kato} and the above observation that
$$ \|J_{high,2}^{3,2}\|_{\infty} \lesssim
\int^t_{t-A} ds \int^{s \wedge (t-\epsilon)}_{s-A} d\tau
\int_{\tau-A-B}^{\tau-\delta}d\alpha \langle t-s\rangle^{-n/2}
\langle s-\tau \rangle^{-n/2} \frac {BR}{\sqrt{N}}
\|U(\alpha)\psi_0\|_{\infty} \lesssim \frac {C_0}{100}t^{-n/2}
\trip \psi_0\trip ,
$$
where $\frac 1{\sqrt{N}}$ is chosen to be sufficiently small to
dominate the implicit constant which only depends on $n,V, \vec
v_2$ and $\epsilon,\delta,A,B$.  Therefore, we conclude that
\eqref{eq:boot1} implies \eqref{eq:boot2}, from which
Theorem~\ref{thm:main} follows.

\section{Decay estimate of the derivatives of $U(t)$}

In this section we prove Theorem~\ref{thm:main2} by induction on
$\kappa$ by following the same scheme of the proof of
Theorem~\ref{thm:main}. The first step is to set up the
cancellation lemma for $\partial U(t)\psi_0$.

\begin{lemma}
\label{lemma:canc4} Let $\kappa$ be a nonnegative integer. Assume
$\sup_{0 \leq \beta \leq \kappa} \sup_{r\in R} \| \widehat
{\partial^\beta V(r)} \|_{L^1} < M $. Let $\alpha$ be a
nonnegative integer n-tuple with $|\alpha|=\kappa$. Suppose $U(t)
$ is the evolution operator of \eqref{eq:2pot} as before. Then
\begin{equation}
\sup_{r \in \R} \|e^{-i(t-s)H_0}V(r) \partial ^\alpha
U(s)\psi_0\|_p < |t|^{-\gamma} M e^{(\kappa +1) Ms}\| \psi_0
\|_{W^{\kappa,p'}} \label{eq:canc4}
\end{equation}
\noindent where $\gamma = n(\frac 12 - \frac 1p)$ and $2 \leq p <
\infty$, $\frac 1p + \frac 1{p'}=1 .$
\end{lemma}

\begin{proof}
Write the left-hand side of \eqref{eq:canc4}$:= \Psi (t,s)$. When
$\kappa =0$, \eqref{eq:canc4} is just the
inequality~\eqref{eq:canc2'}. Note that the
inequality~\eqref{eq:canc2'} holds with $V$ replaced by its
derivative $\partial^\beta V$, as long as $\widehat
{\partial^\beta V} $ lies in $L^1(R^n)$. Assume $\kappa=1$ and
apply Duhamel's formula:

\begin{align*}
& \qquad \|e^{-i(t-s)H_0}V(r) \partial U(s)\psi_0\|_p \\
&\leq \|e^{-i(t-s)H_0}V(r)\partial e^{-isH_0} \psi_0\|_p +
  \int^s_0 \|e^{-i(t-s)H_0}V(r)e^{-i(s- \tau) H_0} \partial V(\tau)U(\tau)\psi_0  \|_p d\tau \\
 & \leq C \|\widehat{V}(r) \|_1 t^{-\gamma} \|\partial \psi_0\|_{p'} +
 \|\widehat{V}(r)\|_1 \int^s_0 \|e^{-i(t-\tau)H_0} (\partial V)(\tau)U(\tau)\psi_0\|_p d
 \tau \\
 &\qquad + \|\widehat{V}(r)\|_1 \int^s_0 \|e^{-i(t-\tau)H_0} V(\tau)
 \partial U(\tau)\psi_0\|_p d\tau \\
 & \leq CM t^{-\gamma} \|\psi_0\|_{W^{1,p'}}+ M \int^s_0 t^{-\gamma} e^{\tau M}
 d\tau \|\psi_0\|_{p'} + M \int^s_0 \Psi (t,\tau) d \tau \\
& \leq CM t^{-\gamma}e^{sM} \|\psi_0\|_{W^{1,p'}} + M \int^s_0
\Psi (t,\tau) d \tau.
\end{align*}
Taking supremum over $r$, we get  $\Psi(t,s)\leq CM
t^{-\gamma}e^{sM}\|\psi_0\|_{W^{1,p'}} + M \int^s_0 \Psi (t,\tau)d
\tau $. By Gronwall's inequality, $\Psi(t,s)\leq CM t^{-\gamma}
e^{2Ms}$.

For $\kappa > 1$, the above argument goes through by induction,
provided that the Fourier transform of the derivatives up to
degree $\kappa$ of $V(r)$ are uniformly bounded in $L^1(\R ^n)$.
\end{proof}

The following is an analog of Corollary~\ref{cor:shorttime}:

\begin{cor} \label{cor:shorttime2}
With the same notations and assumptions as in
Lemma~\ref{lemma:canc4}, we have
\begin{equation}
 \|U(t)\psi_0\|_{W^{\kappa,p}} \lesssim t^{-\gamma} e^{(1+\kappa) Mt}
\|\psi_0\|_{W^{\kappa,p'}}.
 \label{eq:shorttime2}
\end{equation}
\end{cor}

\begin{proof} By Duhamel's formula, Lemma~\ref{lemma:canc4} and the fact that $\partial$ commutes with $
e^{-itH_0}$, we have the following estimate:
\begin{align*}
&\|\partial^\alpha U(t)\psi_0\|_p \les \|e^{-itH_0}
\partial^\alpha \psi_0\|_p + \Sigma_{\beta \leq \alpha} \int^t_0
\|e^{-i(t- \tau) H_0} (\partial ^\beta V)(\tau) \partial
^{\alpha-\beta}U(\tau)\psi_0\|_p d\tau \\
& \lesssim t^{-\gamma} \|\psi_0\|_{W^{\kappa,p'}} + \Sigma_{\beta
\leq \alpha}\int^t_0 t^{-\gamma} e^{(|\beta|+1)M \tau}d\tau
\|\psi_0\|_{W^{\kappa,p'}} \\
& \leq C t^{-\gamma} e^{(\kappa+1)Mt}\|\psi_0\|_{W^{\kappa,p'}}.
\end{align*}
\end{proof}

Similarly, the following lemma generalizes
Lemma~\ref{lemma:canc3}:

\begin{lemma} \label{lemma:canc5} Let $\alpha$ be an n-tuple with
$|\alpha|= \kappa$ and $U(t) $ be the evolution operator of
\eqref{eq:2pot}. For each $m \geq 1$ and $\epsilon >0$, $u_1,
u_2,\ldots,u_m$ are all in either $\R_+$ or $\R_-$, satisfying
$|\sum_{j=1}^{m}u_j|> \epsilon$, then there exists constant
$C=C(m,\epsilon,\kappa,p)$ such that
\begin{equation}
 \|\prod_{j=1}^{m-1} (e^{iu_j H_0}V(s_j))\partial^{\alpha}
U(u_m)\psi_0\|_p \leq  CM^{m-1}  \prod_{j=1}^m \langle u_j
\rangle^{-\gamma} e^{(\kappa+1) M u_m}\|\psi_0\|_{W^{\kappa,p'}},
\label{eq:canc5}
\end{equation}
\noindent where $s_j$ is any real number, $\frac{2n}{n-2} < p <
\infty$, $\frac 1p + \frac 1{p'}=1$ and $$M= \Sigma_{0\leq
\beta\leq \alpha} \sup_{s \in R}(\| \partial^{\beta} V(s)\|_1 +
\|\widehat{\partial^{\beta}V}(s)\|_1) .$$
\end{lemma}

Using Lemma~\ref{lemma:canc4} and Corollary~\ref{cor:shorttime2},
the proof of Lemma~\ref{lemma:canc5} is exactly the same as that
of Lemma~\ref{lemma:canc3}.

We only prove Theorem~\ref{thm:main2} for the case $\kappa =1,2$.
The case $\kappa>2$ can be proved by induction. Specifically, we
prove the following implication:

For any fixed sufficiently large time $T$,

\be \| U(t) \psi_0\|_{W^{\kappa ,p}} \leq C_0  |t|^{-\gamma}
(\|\psi_0\|_{W^{\kappa ,p'}} + e^{-\frac {\alpha T}2}
\|\psi_0\|_{L^2}) \text{\ \ for \ \ }0\le t\le T, \kappa =1,2
\label{eq:boot1'} \ee implies that

\begin{equation}
\|U(t) \psi_0\|_{W^{\kappa,p}} \leq \frac{C_0}{2} |t|^{-\gamma}
(\|\psi_0\|_{W^{\kappa,p'}} + e^{-\frac {\alpha T}2}
\|\psi_0\|_{L^2}) \text{\ \ for\ \ }0\le t\le T, \kappa =1,2
\label{eq:boot2'}
\end{equation}
provided that $\frac {C_0}2$ remains larger than some constant
that does not depend on $T$. The assumption~\eqref{eq:boot1'} can
be made to hold for some $C_0$ depending on~$T$, because of
Corollary~\ref{cor:shorttime2}. Letting $T \to +\infty$ to
eliminate $\|\psi_0\|_{L^2}$, Theorem~\ref{thm:main2} follows from
the iteration of the above implication.

We will first prove \eqref{eq:boot2'} for $\kappa =1 $. For
technical reasons (see \eqref{eq:useboot}), we need the above
bootstrap assumption \eqref{eq:boot1'} for $\kappa +2$. To
simplify the notation, we write $\partial ^ {\alpha}=\partial$ and
$$\|\psi_0\|_{W^{1,p'}} + e^{-\frac {\alpha T}2}
\|\psi_0\|_{L^2}:= \trip \psi_0\trip _{(1,p')}.$$

With these cancellation lemmas for $\partial U(t)\psi_0$, the
proof of Theorem~\ref{thm:main2} follows the scheme of that of
Theorem~\ref{thm:main}. The difference is that now we need to
commute $\partial_x$ with operators such as $e^{itH_0}, \, V$ and
$e^{itH_1}$ to apply the cancellation lemma and the bootstrap
assumption.

We proceed by expanding $U(t)$ with Duhamel's formula:

\begin{align}   & \partial U(t) \psi_0 = \partial e^{-it H_0} \psi_0 - i
\int^t_0 \partial e^{-i(t-s) H_0}V(s) U(s) \psi_0 \, ds \nn \\ & =
\partial e^{-it H_0} \psi_0 - i \int^t_0 e^{-i(t-s) H_0}(\partial V)(s) U(s)
\psi_0 \, ds \nn
\\ & \mbox{\hspace{.3in}}- i \int^t_0 e^{-i(t-s) H_0}V(s) \partial U(s)
\psi_0 \, ds.  \label{eq:DU3}
\end{align}

Notice that $ [\partial, V ]=(\partial V)\cdot$ is a
multiplication operator, which can be viewed as another potential
and Theorem~\ref{thm:main} can be applied to the second term of
\eqref{eq:DU3}. This idea has appeared in the proof of
Lemma~\ref{lemma:canc4}. Specifically, it follows from the proof
of Theorem~\ref{thm:main} and an interpolation with the $L^2$
conservation of $U(t)$ that

$$ \|\int^t_0 e^{-i(t-s) H_0} V(s) U(s) \psi_0 \, ds\|_p \lesssim
t^{-\gamma} \|\psi_0\|_{p'}.
$$

By assumption, $\partial V_j $ satisfies the regularity and
smoothness conditions for $V_j$ in Theorem~\ref{thm:main}, and we
conclude that

$$ \|\int^t_0 e^{-i(t-s) H_0} (\partial V)(s) U(s) \psi_0 \, ds\|_p \lesssim
t^{-\gamma} \|\psi_0\|_{p'}.
$$

We expand the last term of \eqref{eq:DU3} by Duhamel's formula
just as in Section~\ref{sec:decay} and perform the same
decomposition. With the cancellation lemma for $\partial U(t)$ and
Remark~\ref{rem:bdsates}, the last term~\eqref{eq:DU3} is reduced
to the following:

\be  \label{eq:hard2'} \sum_{j=1}^2 \int^t_{t-A} ds \int^{s \wedge
(t-\epsilon)}_{s-A} d\tau \, e^{-i(t-s) H_0}V(s) e^{-i(s-\tau)
H_0}V_j(\cdot-\tau \vec v_j ) \partial P_c(H_1,\tau)U(\tau)
\psi_0. \ee

Before we proceed, we observe that our assumptions guarantee

\be \|P_{c}(H_1) e^{-it H_1} \psi_{0}\|_{L^q}\leq C_q\,
|t|^{-\gamma}\|\psi_{0}\|_{L^{q'}}.  \label{eq:interpo1} \ee This
implies

\begin{align*}  &\|H_1 P_{c}(H_1) e^{-it H_1}  \psi_{0}\|_{L^q}= \|
P_{c}(H_1) e^{-it H_1} H_1 \psi_{0}\|_{L^q} \\
& \leq C_q\, |t|^{-\gamma}\| H_1 \psi_{0}\|_{L^{q'}} \leq C_q \,
|t|^{-\gamma}\| \psi_{0}\|_{W^{2,q'}}. \end{align*}

As $V_1 \in L^{\infty}(\R^n)$ and double Riesz transforms is
bounded  on $L^q(\R^n) \,\, 1< q < +\infty$, the above inequality
in the case of $ 1< q < +\infty$, implies that

\be \|P_{c}(H_1) e^{-it H_1}  \psi_{0}\|_{W^{2,q}} \lesssim
|t|^{-\gamma}\| \psi_{0}\|_{W^{2,q'}}. \label{eq:interpo2} \ee

Interpolating between \eqref{eq:interpo1} and \eqref{eq:interpo2}
(Theorem 6.4.5 \cite{BL}), we conclude that

\begin{equation} \label{eq:inter1} \|P_{c}(H_1) e^{-it H_1}
\psi_{0}\|_{W^{1,q}}\leq C_q\,
|t|^{-\gamma}\|\psi_{0}\|_{W^{1,q'}}.
\end{equation}

\noindent where $2 \leq q < \infty$, $\frac 1q + \frac 1{q'}=1$
and $\gamma= n(\frac 12- \frac 1q)$. Because double Riesz
transforms are unbounded  on $L^\infty (\R^n)$, we exclude $p =
\infty$ in Theorem~\ref{thm:main2}.

We write $P_c(H_1)U(\tau)= P_c(H_1)e^{-i\tau H_1}-i P_c(H_1)
\int_0^{\tau} e^{-i(\tau-r) H_1}V_2(r)U(r)\, dr$ and
\eqref{eq:hard2'} is broken into two terms.

It follows from \eqref{eq:inter1}, among other things that the
first term of \eqref{eq:hard2'}, which contains $P_c(H_1)e^{-i\tau
H_1}$ is dominated by $|t|^{-\gamma}\|\psi_{0}\|_{W^{1,p}}$.

The second term of \eqref{eq:hard2'} is decomposed as follows:

\be \int_0^{\tau}dr = \int_0^{\delta}dr +\int_ {\delta}^{A}
dr+\int_{A}^{\tau-A}dr +\int_{\tau-A}^{\tau-\delta}dr +
\int_{\tau-\delta}^{\tau}dr. \label{eq:decomp'} \ee

We estimate each term in \eqref{eq:decomp'} with similar methods
as that for \eqref{eq:decomp2}. Because of \eqref{eq:inter1}, the
terms containing $\int_ {\delta}^{A} dr$ and $\int_{A}^{\tau-A}dr$
in \eqref{eq:decomp'} can be estimated exactly as that there is no
derivative before $P(H_1)$, and we omit the details here. Again by
\eqref{eq:inter1} with $q=2$, the term containing
$\int_{\tau-\delta}^{\tau}dr$ in \eqref{eq:decomp'} is estimated
as follows:

\begin{align*}\int^t_{t-A} ds \int^{s \wedge (t-\epsilon)}_{s-A} &
 d\tau \int_{\tau- \delta}^{\tau} dr
\|e^{-i(t-s) H_0}V(s) e^{-i(s-\tau) H_0}V_1(\tau)
\partial e^{-i(\tau-r) H_1}P_c(H_1) V_2(r)U(r) \psi_0\|_p \\
& \lesssim  \sup _{t-2A< \tau <t} \int_{\tau-\delta}^\tau dr
\|\partial P_c(H_1)e^{-i(\tau-r) H_1} V_2(r)U(r)\psi_0\|_2\\
& \lesssim  \sup _{t-2A< \tau <t} \int_{\tau-\delta}^\tau dr
\| V_2(r)U(r)\psi_0\|_{W_{1,2}}\\
& \lesssim  \sup _{t-2A< \tau <t} \int_{\tau-\delta}^\tau dr
\|U(r)\psi_0\|_{W_{1,p}}\\
& \lesssim t^{-\gamma} C_0 \delta \trip \psi_0\trip _{(1,p')} \leq
\frac {C_0}{100} t^{-\gamma} \trip \psi_0\trip _{(1,p')}.
\end{align*}
Here $\delta >0$ is chosen sufficiently small.

The $\int_0^{\delta}dr$ term in \eqref{eq:decomp'} is expanded
 by Duhamel's formula: $$e^{-i(\tau-r) H_1} = e^{-i(\tau-r) H_0}-i\int_0^{\tau-r}
 e^{-i(\tau-r-\beta) H_1}V_1 e^{-i\beta H_0} \, d\beta.$$

Plugging the above expression into the $\int_0^{\delta}\, dr$
term, we get two terms. The first one containing $ e^{-i(\tau-r)
H_0}$ is

\be \int^t_{t-A} ds \int^{s \wedge (t-\epsilon)}_{s-A} d\tau
\int_0^{\delta} dr e^{-i(t-s) H_0}V(s) e^{-i(s-\tau) H_0}V_1(\tau)
 \partial P_c(H_1) e^{-i(\tau-r) H_0}V_2(r)U(r) \psi_0 .  \label{eq:hard31'}\ee

Since $P_c(H_1)= Id - P_b(H_1)$ and $P_b(H_1)$ is a bounded
operator from $L^p$ to $L^p$, $P_c(H_1)$ is bounded from $L^p$ to
$L^p$. It follows from Lemma~\ref{lemma:canc4}, $0<r< \delta$, and
the Leibnitz rule that

\begin{align*} &\|H_1 P_c(H_1) e^{-i(\tau-r)
H_0}V_2(r)U(r) \psi_0 \|_p =\|P_c(H_1) H_1  e^{-i(\tau-r)
H_0}V_2(r)U(r) \psi_0 \|_p \\
& \leq C \|H_1  e^{-i(\tau-r) H_0}V_2(r)U(r) \psi_0 \|_p \\
& \leq C \|V_1\|_\infty \|e^{-i(\tau-r) H_0}V_2(r)U(r) \psi_0 \|_p + \|e^{-i(\tau-r) H_0} \Delta V_2(r)U(r) \psi_0 \|_p\\
& \leq C \tau ^{-\gamma} \|\psi_0\|_{W^{2,p'}}.
\end{align*}
Since $H_1=H_0 + V_1$ and $V_1$ is bounded, we see that
$$
\| \Delta P_c(H_1) e^{-i(\tau-r) H_0}V_2(r)U(r) \psi_0 \|_p \leq C
\tau ^{-\gamma} \|\psi_0\|_{W^{2,p'}}.
$$
Because the double Riesz transforms are bounded on $L^p(\R^n)\,\,
1<p<\infty$, it follows that
$$
\|P_c(H_1) e^{-i(\tau-r) H_0}V_2(r)U(r) \psi_0 \|_{W^{2,p}} \leq C
\tau ^{-\gamma} \|\psi_0\|_{W^{2,p'}}.
$$
Therefore, by complex interpolation, we see \be \label{eq:inter2}
\|P_c(H_1) e^{-i(\tau-r) H_0}V_2(r)U(r) \psi_0 \|_{W^{1,p}} \leq C
\tau ^{-\gamma} \|\psi_0\|_{W^{1,p'}}. \ee

\noindent which implies that $\|\eqref{eq:hard31'}\|_{W^{1,p}}
\lesssim t^{-\gamma} \|\psi_0\|_{W^{1,p'}}$.

For the term containing $\int_0^{\delta} dr \int_0^{\tau-r}d\beta$
, we perform the exact same decomposition as in \eqref{eq:hard32}
and each step there goes through provided \eqref{eq:inter1} and
\eqref{eq:inter2}.

The term containing  $\int_{\tau - A}^{\tau-\delta} dr $  in
\eqref{eq:decomp'} is

\be  \label{eq:hard3'}  \int^t_{t-A} ds \int^{s \wedge
(t-\epsilon)}_{s-A} d\tau \int_{\tau-A}^{\tau-\delta}dr \,
\|e^{-i(t-s)H_0}V(s) e^{-i(s-\tau)H_0}V_1 \partial P_c(H_1)
e^{-i(\tau-r) H_1}V_2(r)U(r) \psi_0 \|_p . \ee

The proof of Theorem~\ref{thm:main} showed that $\forall \epsilon
>0$, the following holds:

$$ \|V_1 P_c(H_1) e^{-i(\tau-r) H_1}V_2(r)U(r)
\psi_0\|_{\infty} < \epsilon C_0 t^{-\frac n2}\|\psi_0\|_1,
$$
given $t$ sufficiently large. Going through the proof, we see that
the same argument also shows

$$ \|V_1 P_c(H_1) e^{-i(\tau-r) H_1}V_2(r)U(r)
\psi_0\|_p < \epsilon C_0 t^{-\gamma}\|\psi_0\|_{p'}.
$$

Furthermore the above inequality holds if $V_1$ or $V_2$ is
replaced by its derivative. Another observation is that, given our
new cancellation lemma for $\partial U(r) \psi_0$,

$$ \|V_1 P_c(H_1) e^{-i(\tau-r) H_1}V_2(r)\partial^\beta U(r)
\psi_0\|_p < \epsilon C_0 t^{-\gamma}\trip \psi_0\trip
_{(|\beta|,p')}.
$$
Indeed, to prove the above inequality, we decompose the left-hand
side into a high velocity part and a low velocity part. Each part
generates the small constant $\epsilon$
 for the same reason as in Section $3.3$. The same argument with
 the bootstrap assumption \eqref{eq:boot1'} implies:

\begin{align} & \|V_1 H_1 P_c(H_1) e^{-i(\tau-r)
H_1}V_2(r)U(r) \psi_0\|_p  \nn \\
 & = \|V_1 P_c(H_1) e^{-i(\tau-r) H_1} H_1 V_2(r)U(r) \psi_0\|_p
  \lesssim \epsilon C_0 t^{-\gamma}\trip \psi_0\trip _{(2,p')}.
  \label{eq:useboot}
\end{align}

It follows from the above inequality and an elementary calculation
that

 \be  \| V_1 P_c(H_1) e^{-i(\tau-r) H_1}V_2(r)U(r)
\psi_0\|_{W^{2,p}} \lesssim \epsilon C_0 t^{-\gamma}|\! \|\psi_0
|\!\|_{(2,p')} \ee

Hence, by complex interpolation, for $\forall \, \epsilon
>0$,

\be \|V_1 P_c(H_1) e^{-i(\tau-r) H_1}V_2(r)U(r) \psi_0\|_{W^{1,p}}
 \lesssim \epsilon C_0 t^{-\gamma}|\!\|\psi_0|\!\|_{(1,p')}. \ee
given $t$ sufficiently large. This implies that
$\|\eqref{eq:hard3'}\|_{W^{1,p}}$ can be estimated by $\frac
1{100} C_0 t^{-\gamma}\trip \psi_0\trip _{1,p'}$.

Therefore, we proved \eqref{eq:boot2'} for $\kappa=1$. The same
procedure also proves \eqref{eq:boot2'} for $\kappa=2$. Thus, we
finish the bootstrap argument and conclude that

$$\|U(t)\psi_0\|_{W^{\kappa,p}} \lesssim
\|\psi_0\|_{W^{(\kappa,p')}},$$

\noindent by letting $T \to \infty$. The proof for $\kappa
>2$ is similar by induction. Thus we have proved Theorem~\ref{thm:main2}.

\section{Boundedness of the Sobolev norm of $U(t,s)\psi_0 $ }

The goal of this section is to prove Theorem~\ref{thm:main3} when
$\kappa$ is a positive integer. The intuition comes from the case
$ \kappa =1$ (\cite{Gr}). To bound the kinetic energy (the $H^1$
norm), we look at the observable $K(t)= \frac 12(p-\frac xt)^2+
\sum_{l=1}^{m}V_l(t)$. $\la K(t) \ra$ will decrease if the
particle is far away from any potential, since the observable
$(p-\frac xt)^2 $ decreases like $t^{-2}$ for the free motion (the
Pseudo-conformal identity). If the particle is close to the center
of potential $V_l$, then $\frac xt \approx \vec {v_l} $ and $\la
K(t) \ra \approx \la \frac 12(p-\vec {v_l})^2+ V_l(x- \vec {v_l}
t) \ra $, which clearly is the total energy of this one potential
stationary subsystem up to a Galilean transform. To carry this
boundedness from $\la K(t) \ra$ to $\la p^2 \ra $, we need to
replace the vector field $\frac xt$ by $ \nu(x,t)$, such that
$\nu(x,t)$ is uniformly bounded and is equal to $\vec {v_l} $ in
an increasingly big neighborhood of $x=\vec {v_l} t $.

Vigorously, consider a smooth, uniformly bounded vector field
$$\nu(x,t): \R^n \times (-\infty,-T]\cup [T,+\infty) \to \R^n $$ and
let $$ K_0(t)= \frac 12(p-\nu(x,t))^2+ \sum_{l=1}^{m}V_l(t) ,
$$ where $T$ is a large positive constant, $p=(p_1,\cdots, p_n)$ and
$p_j=-i \frac {\partial}{\partial x_j}$. Note $p^2=H_0$ and $\frac
12(p-\nu(x,t))^2$ is a well-defined self-adjoint positive
operator.

In \cite{Gr}, Graf constructed $\nu(x,t) $ and proved
$\|U(t,s)\psi_0\|_{H^1}$ is bounded as $t \to \infty$ by bounding
$\frac d{dt}\langle K_0(t) \rangle$ from above by a
time-integrable function, where $\langle K_0(t)
\rangle=(U(t,s)\psi_0, K_0(t)U(t,s)\psi_0)_{L^2}$. We write
$(f,g)$ as the inner product of $f,\, g$ in the $L^2(\R)$ sense.

To prove Theorem~\ref{thm:main3}, we need to define the proper
analog of $K_0(t)$ suitable to the $H^\kappa$ norm of
$U(t,s)\psi_0$ to match the intuition given by the classical
system. Fortunately the following observable works:

$$K(t)= \sum_{l=1}^m (\frac 12(p-\nu(x,t))^2+V_l(t))^{\kappa} -
(m-1)(\frac 12(p-\nu(x,t))^2)^{\kappa}.$$ Notice that
$K(t)=K_0(t)$ if $\kappa =1$. Because $\nu(x,t)$ and its
derivatives are bounded uniformly in space time and $V_j \in
C_0^{\kappa}(\R^n)$, we have the following, writing $\langle
K(t,s) \rangle = (U(t,s)\psi_0, K(t)U(t,s)\psi_0)$:

$$\|U(t,s)\psi_0\|^2_{H^\kappa} \lesssim \langle K(t,s) \rangle +
\|U(t,s)\psi_0\|^2_{H^{\kappa-1}}; \quad \langle K(t,s) \rangle
\lesssim \|U(t,s)\psi_0\|^2_{H^\kappa}.$$

By induction on $\kappa$, it suffices to show $\langle K(t,s)
\rangle$ is bounded uniformly in $t$ and $s$.

Expand $K(t)$ as polynomial of $(\frac 12(p-\nu(x,t))^2$. Though
$(\frac 12(p-\nu(x,t))^2$ and $V_l(t)$ do not commute with each
other, viewing $K(t)$ as a differential operator, the term of
highest degree is $((\frac 12(p-\nu(x,t))^2)^\kappa$ which is
positive, self-adjoint. The other terms in $K(t)$ are of degree no
bigger than $2\kappa -2$ with bounded and smooth enough
coefficients. Correspondingly, $\langle K(t,s) \rangle$ breaks
into two parts. The part $(U(t,s)\psi_0 , (\frac
12(p-\nu(x,t))^2)^\kappa U(t,s)\psi_0 )$ is always nonnegative.
The other part containing the low degree terms can be dominated by
$\|U(t,s)\psi_0\|^2_{H^{\kappa-1}}$. By the induction hypothesis,
it follows that $\langle K(t,s) \rangle$ is bounded from below. To
bound  $\langle K(t,s) \rangle$ from above, it suffices to show
that for $t>T$

\be \label{eq:gronwell}\frac d{dt}\langle K(t,s) \rangle \leq
t^{-(1+\delta)}C(\langle K(t,s)\rangle +
\|U(t,s)\psi_0\|^2_{H^{\kappa -1}}).\ee

For $t<-T$, the opposite of the above inequality should hold:

\be \label{eq:gronwell'}\frac d{dt}\langle K(t,s) \rangle \geq
t^{-(1+\delta)}C(\langle K(t,s)\rangle +
\|U(t,s)\psi_0\|^2_{H^{\kappa -1}}).\ee

First let's consider $t>T$, integrating \eqref{eq:gronwell},

$$\langle K(t_2,s)\rangle-\langle K(t_1,s)\rangle \leq C
\int_{t_1}^{t_2} t^{-(1+\delta)} \langle K(t,s)\rangle dt + C
\sup_{t,s \in \R} \|U(t,s)\psi_0\|^2_{H^{\kappa -1}}.$$ Choosing
$T
>0$ large enough such that $C \int_T^\infty t ^{-1-\delta}dt < \frac
12$, then

$$ \langle K(t_2,s)\rangle \leq \langle K(t_1,s)\rangle + C
\sup_{t,s \in \R} \|U(t,s)\psi_0\|^2_{H^{\kappa -1}}+ \frac 12
\max_{t_1<t<t_2} \langle K(t,s)\rangle $$

By \eqref{eq:compact}, $\max_{t_1<t<t_2} \langle K(t,s)\rangle <
\infty $. This implies that

$$ \max_{t_1<t<t_2} \langle K(t,s)\rangle \leq 2 \langle K(t_1,s)\rangle + C
\|\psi_0\|^2_{H^{\kappa -1}}$$

Letting $t_2 \to +\infty$ and $t_1=T$, it follows that $
\max_{t>T} \langle K(t,s)\rangle < C \langle K(T,s)\rangle + C
\|\psi_0\|^2_{H^{\kappa -1}}$. Hence $\langle K(t,s)\rangle \leq
C_T \|\psi_0\|_{H^{\kappa}}$ for $t
> T$ and $s \in [-T,T]$. For $t<-T$, we integrate \eqref{eq:gronwell'} to
bound $\langle K(t,s) \rangle$
from above and Theorem~\ref{thm:main3} follows in this case by the
same argument given that \eqref{eq:gronwell'} holds.

Before we proceed to prove \eqref{eq:gronwell} and
\eqref{eq:gronwell'}, let's specify some properties of the vector
field $\nu(x,t)$. It is convenient to describe $\nu(x,t)$ in the
rescaled coordinates $y=\frac xt$. Let $u_0=2 \max_{1 \leq l \leq
m} |\vec{v_l}|$. When $|y|> u_0$, $\nu(x,t)=u_0 \frac y{|y|}$.
When $y \in B_l$, we specify $\nu(x,t)=\vec{v_l}$, where $B_l$ is
a fixed ball centered at $\vec{v_l}$. We suppose that $B_l \,\,
(l=1,\cdots, m)$ lie in the big ball $B_0$ centered at the origin
with radius $u_0$ and that they are disjoint from each other. When
$y \in B_0- \cup_{l=1}^m B_l $, we specify $\nu(x,t)=y$. To make
the vector field smooth, we modify and smooth the vector field in
the scale of $|t|^{1-\gamma}$, where $\gamma$ is a small positive
number. In the rescaled coordinates $y$, the scale is
$|t|^{-\gamma}$. Specifically, consider

$$\omega(s,\alpha)= s\, \varphi (\frac {u_0-s}\alpha ) + u_0 (1- \varphi
(\frac {u_0-s}\alpha )),$$ where $\varphi \in C^\infty(\R) $ with
$\varphi' \geq 0 $ and
$$\varphi(x)=0 \,\, \text{for}\,\, x\leq 0 \qquad \varphi(x)=1 \,\, \text{for}\,\, x>1 .$$

Then writing $y=\frac xt$, we define
$$\omega^{(0)}(x,t)= \omega (|y|, |t|^{-\gamma}) \frac y{|y|} \quad \text{
and} \quad \omega^{(\ell)}(x,t)=-(y- \vec{v_\ell}) \varphi (2-
|t|^\delta |y-\vec{v_\ell}|), $$

\noindent where $\ell =1,2,\cdots,m$. Finally,  $\nu(x,t):=
\sum_{\ell= 0}^m \omega^{(\ell)}$

The properties of the vector field $\nu(x,t) $ that concern us are
listed as follows:

\begin{enumerate}
\item $\nu$ is bounded in space time. The $k$-th space derivatives
of $\nu$ uniformly decay as $ |t|^{-k(1-\gamma)}$ as $t \to
\infty$.

\item $(\nu_{i,j})_{n \times n}$ as a matrix is symmetric and
positive semi-definite when $t>0$, negative semi-definite when
$t<0$, where $\nu_i$ is the $i$-th component of vector $\nu$ and
the indices following a comma stand for partial derivatives in
space. As $\nu_{k,j}=\nu_{j,k}$, $p_k-\nu_k$ and $p_j-\nu_j$
commute with each other, i.e. $[p_k-\nu_k,p_j-\nu_j]=0$.

\item $\|\nu_{i,j}\nu_ j + \frac {\partial \nu_i}{\partial
t}\|_{\infty} \leq C |t|^{-(1+\delta)}$. Here we make the choice
$1+\delta = \min \{ 1+ \gamma, 2-2\gamma \} >1 $. Summation over
double indices is understood.
\end{enumerate}

These properties can be shown by a direct calculation (\cite{Gr}).
 Now we are going to prove
\eqref{eq:gronwell} and \eqref{eq:gronwell'} and proceed by
observing that
\begin{align}
& i\frac {\partial}{\partial t}U(t,s)\psi_0= H(t) U(t,s)\psi_0 \, \quad \textrm{ and } \\
& -i\frac {\partial}{\partial s}U(t,s)\psi_0=  U(t,s) H(s)\psi_0 .
\end{align}

It follows from the above that $\frac d{dt}\langle K(t,s) \rangle
= (U(t,s)\psi_0,(i[ H(t),K(t)] +\frac{\partial K}{\partial t})
U(t,s)\psi_0)$. A straightforward calculation shows:
\begin{align*} &\frac{\partial K}{\partial t} = \sum_{l=1}^m
\sum_{k=0}^{\kappa -1} (\frac 12 (p-\nu(x,t))^2+V_l(t))^k \frac
d{dt}(\frac 12 (p-\nu)^2+V_l(t)) (\frac 12 (p-\nu)^2+V_l(t))^{\kappa-1-k} \\
& \quad -(m-1)\sum_{k=0}^{\kappa -1} (\frac 12 (p-\nu)^2)^k \frac
d{dt} \frac 12 (p-\nu(x,t))^2 (\frac 12 (p-\nu)^2)^{\kappa-1-k} :=
J_1+J_2,
\end{align*} and the commutator

$$ [ H(t),K(t) ]= [\frac 12 p^2 + \sum_{l=1}^m V_l(t), \sum_{l=1}^m
(\frac 12(p-\nu)^2+V_l(t))^{\kappa} -(m-1)(\frac
12(p-\nu)^2)^{\kappa} ]$$

\begin{align*}
&= \sum_{l=1}^m \sum_{k=0}^{\kappa -1}(\frac 12
(p-\nu)^2+V_l(t))^k [\frac 12 p^2, \frac 12 (p-\nu)^2+V_l(t) ]
(\frac12(p-\nu)^2+V_l(t))^{\kappa-1-k} \\
&-(m-1)\sum_{k=0}^{\kappa -1} (\frac 12 (p-\nu)^2)^k  [ \frac 12
p^2, \frac 12 (p-\nu)^2  ] (\frac 12 (p-\nu)^2)^{\kappa-1-k}  \\
&+ \sum_{l,j=1}^m \sum_{k=0}^{\kappa -1}(\frac 12
(p-\nu)^2+V_l(t))^k  [V_j(t), \frac 12 (p-\nu)^2+V_l(t)]
(\frac 12 (p-\nu)^2+V_l(t))^{\kappa-1-k} \\
&-(m-1)\sum_{j=1}^m \sum_{k=0}^{\kappa -1} (\frac 12 (p-\nu)^2)^k
 [ V_j(t), \frac 12 (p-\nu)^2 ] (\frac 12 (p-\nu)^2)^{\kappa-1-k}
:= J_3+ J_4 + J_5 + J_6.
\end{align*}

First, let's consider \be J_1+i J_3 + i J_5 =\sum_{l=1}^m
\sum_{k=0}^{\kappa -1}(\frac 12 (p-\nu)^2+V_l(t))^k M_1 (\frac
12(p-\nu)^2+V_l(t))^{\kappa-1-k}, \label{eq:middle1}  \ee where $
M_1=i[\frac12 p^2 + \sum_{j=1}^m V_j(t), \frac12(p-\nu)^2+V_l(t)]
+ \frac d{dt}(\frac 12 (p-\nu)^2+V_l(t)). $ Another elementary
calculation gives the following:

\begin{align} & M_1= i [\sum_{j\neq l}V_j , \frac 12 (p-\nu)^2+V_l ]+ A -
\frac 12 p_i (\nu_{i,j}\nu_ j + \frac {\partial \nu_i}{\partial
t}) -\frac 12 (\nu_{i,j}\nu_ j+ \frac {\partial \nu_i}{\partial
t})p_i  \nn  \\
& \qquad \qquad +\nu_i (\nu_{i,j}\nu_ j + \frac {\partial
\nu_i}{\partial t}) + \frac 14 \nu_{i,ijj} + (\nu -
\vec{v_l})\cdot \nabla V_l , \label{eq:middle}
\end{align}

\noindent where $A=-(p_i-\nu_i) \frac {\nu_{i,j}+\nu_{j,i}}2
(p_j-\nu_j)$ is a symmetric, semi-definite negative operator when
$t> 0$ and semi-definite positive operator when $t< 0$. It follows
from the properties of $ \nu_{x,t}$ that

\be \label{eq:junk1} \|\nu_{i,j}\nu_ j + \frac {\partial
\nu_i}{\partial t}\|_{\infty} \leq C |t|^{-(1+\delta)}, \quad
\|\nu_{i,ijj}\|_\infty \leq C |t|^{-1-\delta} \ee

\noindent and that the $L_\infty$ norm of derivatives of these
terms decay even faster because each space derivative gains a
factor $|t|^{\delta -1}$. Moreover $\sum_{l=1}^m (\nu -
\vec{v_l})\cdot \nabla V_l$ vanishes as $|t|>T$ is sufficiently
large, since $\nu - \vec{v_l}$ vanishes on an increasing
neighborhood of $x=t \vec{v_l}$, which will eventually contain the
support of $\nabla V_l$.

Plugging the expression of $M_1$ into
expression~\eqref{eq:middle1}, we claim the decaying terms listed
in equation~\eqref{eq:junk1} only produce time integrable term. We
calculate the term containing $\frac 12 p_i (\nu_{i,j}\nu_ j +
\frac {\partial \nu_i}{\partial t})$ as an example to illustrate
this point:

\begin{align} &|(U(t,s)\psi_0, (\frac 12 (p-\nu)^2+V_l(t))^k \frac 12 p_i
(\nu_{i,j}\nu_ j + \frac {\partial \nu_i}{\partial t})(\frac 12
(p-\nu)^2+V_l(t))^{\kappa-1-k}U(t,s)\psi_0 )| \nn \\
&=|( \frac 12 p_i(\frac 12 (p-\nu)^2+V_l(t))^k U(t,s)\psi_0,
(\nu_{i,j}\nu_ j + \frac {\partial \nu_i}{\partial t})(\frac 12
(p-\nu)^2 + V_l(t))^{\kappa-1-k} U(t,s)\psi_0 )|.
\label{eq:commutator}
\end{align}

If $2k+1=\kappa$ or $2k+2=\kappa$, \eqref{eq:commutator} can be
dominated by

\begin{align*}& C|t|^{-1-\delta} \|p_i(\frac 12 (p-\nu)^2+V_l(t))^k
U(t,s)\psi_0\|_{L^2} \|(\frac 12 (p-\nu)^2 + V_l(t))^{\kappa -1-k} U(t,s)\psi_0\|_{L^2} \\
& \leq C|t|^{-1-\delta} \|U(t,s)\psi_0\|^2_{H^{\kappa}} \\
& \leq C|t|^{-1-\delta} ( \langle K(t,s)\rangle +
\|U(t,s)\psi_0\|^2_{H^{\kappa -1}} ).
\end{align*}

If $\kappa \neq 2k+1 $ or $2k+2$, first consider $2k+2 < \kappa $
and $\kappa =2d+1$, an odd integer. We need to commute
$\nu_{i,j}\nu_ j + \frac {\partial \nu_i}{\partial t}$ with
$(\frac 12 (p-\nu)^2+V_l)^{d-k}$. Specifically, we claim that

$$(\frac 12 (p-\nu)^2 + V_l(t))^{d-k}(\nu_{i,j}\nu_ j + \frac {\partial \nu_i}{\partial
t}) \frac {p_i}2 (\frac 12 (p-\nu)^2 + V_l(t))^k
$$
is an differential operator of degree $2d+1$, whose coefficients
are of magnitude $t^{-1-\delta}$. This is clear because
$\nu_{i,j}\nu_ j + \frac {\partial \nu_i}{\partial t}$ and its
derivatives decay at least as $|t|^{-1-\delta}$. Hence,
\eqref{eq:commutator} is dominated by $C|t|^{-1-\delta} (\langle
K(t,s)\rangle + \|U(t,s)\psi_0\|^2_{H^{\kappa -1}})$. In the case
that $2k+2 < \kappa $ and $\kappa =2d$ or $2k+1 > \kappa  $,
\eqref{eq:commutator} is dominated by $C|t|^{-1-\delta} (\langle
K(t,s)\rangle + \|U(t,s)\psi_0\|^2_{H^{\kappa -1}})$ due to the
same reason.

Therefore, it remains to estimate the following in
expression~\eqref{eq:middle1} : \be \sum_{l=1}^m
\sum_{k=0}^{\kappa -1}(\frac 12 (p-\nu)^2+V_l(t))^k (i
[\sum_{j\neq l}V_j , \frac 12 (p-\nu)^2+V_l ]+ A) (\frac
12(p-\nu)^2+V_l(t))^{\kappa-1-k}. \label{eq:middle1''}\ee

Observe that for given time $t$, $\nu(x,t)$ is a constant vector
on a ball centered at $t \vec v_{l} $ with radius growing linearly
in $|t|$ approximately. So as long as $|t|$ is large, $\nu(x,t)$
will be constant on the support of $V_l(t)$. This implies that
$\nu_{j,i},\nu_{i,j}$ both vanish on the support of $V_l(t)$.
Hence it follows from $A=-(p_i-\nu_i) \frac {\nu_{i,j}+\nu_{j,i}}2
(p_j-\nu_j)$ that $A \, V_l =0$ and $ V_l \, A=0$. Moreover, for
$j \neq l$, $V_j(t), V_l(t)$ have disjoint supports given that $t$
is large. So the expression~\eqref{eq:middle1''} is reduced to the
following:

\begin{align} & \sum_{l=1}^m \sum_{k=0}^{\kappa -1}(\frac 12 (p-\nu)^2)^k
(i [\sum_{j\neq l}V_j , \frac 12 (p-\nu)^2]+ A)
(\frac 12(p-\nu)^2)^{\kappa-1-k}\\
&=\sum_{k=0}^{\kappa -1}(\frac 12 (p-\nu)^2)^k (m A + (m-1)i
[\sum_ j V_j , \frac 12 (p-\nu)^2]) (\frac
12(p-\nu)^2)^{\kappa-1-k}
 \label{eq:middle1'} \end{align}


Secondly, we consider

\be J_2+iJ_4+iJ_6=-(m-1)\sum_{k=0}^{\kappa -1} (\frac 12
(p-\nu)^2)^k (i[\frac 12 p^2+\sum_{j=1}^m V_j(t), \frac 12
(p-\nu)^2 ]+ \frac d{dt} \frac 12 (p-\nu)^2 )(\frac 12
(p-\nu)^2)^{\kappa-1-k}. \label{eq:middle2} \ee

Setting all potentials $V_l = 0$ in \eqref{eq:middle}, we see that

$$i[\frac 12 p^2, \frac 12 (p-\nu)^2 ]+ \frac d{dt} \frac 12
(p-\nu)^2 = A + \texttt{time integrable terms},
$$
where the time integrable terms equal to $$  A - \frac 12 p_i
(\nu_{i,j}\nu_ j + \frac {\partial \nu_i}{\partial t}) -\frac 12
(\nu_{i,j}\nu_ j+ \frac {\partial \nu_i}{\partial t})p_i +\nu_i
(\nu_{i,j}\nu_ j + \frac {\partial \nu_i}{\partial t}) + \frac 14
\nu_{i,ijj}
$$
and can be estimated exactly as those in $J_1+iJ_3+iJ_5$. We are
left to estimate in $J_2+iJ_4+iJ_6$:

\be -(m-1)\sum_{k=0}^{\kappa -1} (\frac 12 (p-\nu)^2)^k (A+i[
\sum_{j=1}^m V_j(t),\frac 12 (p-\nu)^2 ] )(\frac 12
(p-\nu)^2)^{\kappa-1-k} \label{eq:middle2'} \ee

Now adding \eqref{eq:middle1'} and \eqref{eq:middle2'} together,
we see that $\la i[ H(t),K(t)] +\frac{\partial K}{\partial t} \ra$
is simplified as some time integrable terms plus the following:

\be \sum_{k=0}^{\kappa -1} (\frac 12 (p-\nu)^2)^k A (\frac 12
(p-\nu)^2)^{\kappa-1-k} \label{eq:middle3}, \ee

\noindent which is a differential operator of degree $2\kappa$.

First we observe  that $[p_k-\nu_k, p_j-\nu_j]=0$ and
$(p_k-\nu_k)\frac {\nu_{i,j}+\nu_{j,i}}2 = \frac
{\nu_{i,j}+\nu_{j,i}}2 (p_k-\nu_k) + \frac
{\nu_{i,jk}+\nu_{j,ik}}2$. Second $\nu_{i,jk}+\nu_{j,ik}$ and its
derivatives decay at least as fast as $|t|^{-1-\delta}$ when $ t
\to \infty$ and thus is integrable in time. Hence if we commute
$A$ with $(p-\nu)^2$ or $p_j-\nu_j$, the commutator is time
integrable.

If $\kappa=2d+1$, an odd integer, then $$(\frac 12 (p-\nu)^2)^k A
(\frac 12 (p-\nu)^2)^{\kappa-1-k}= (\frac 12 (p-\nu)^2)^d A (\frac
12 (p-\nu)^2)^d + \texttt {time-integrable terms}.$$ The first
summand is negative (positive) definite when $t>0$ ($t<0$).

If $\kappa=2d$, an even integer, then $(\frac 12 (p-\nu)^2)^k A
(\frac 12 (p-\nu)^2)^{\kappa-1-k}= (\frac 12 (p-\nu)^2)^{h-1}\frac
12 (p_j-\nu_j)A(p_j-\nu_j) (\frac 12 (p-\nu)^2)^{h-1} + \texttt
{time-integrable terms}$. Again the first summand is negative
(positive) definite if $t>0$  ($t<0$).

Hence, we have written $\frac d{dt} \langle K(t,s)\rangle$ as a
sum of a negative (positive if $t<0$) term and other
time-integrable terms. More precisely, the time-integrable terms
decay at least as fast as $|t|^{-1-\delta}$. Therefore, we have
proved \eqref{eq:gronwell} for $t> T$ and \eqref{eq:gronwell'} for
$t<-T$.

Finally, we deal with the case where $|t|<T,\, s > T$ by time
reversal. Write $r=s-t$ and $ \widetilde{U}(r,s)= U(s-r,s)$,
$\widetilde{H}(r)=H(s-r)$. Then we have $i \partial_r
\widetilde{U}(r,s)= - \widetilde{H}(r) \widetilde{U}(r,s)$. Define
the corresponding observable:

$$\widetilde{K}(r)= \sum_{l=1}^m (\frac 12(p+\nu(x,s-r))^2+V_l(x-s \vec
v_l + \vec v_l r))^{\kappa} - (m-1)(\frac 12(p+\nu(x,s-r))^2)
^{\kappa}.$$

It can be shown that $\widetilde{U}(r,s)$ is a bounded operator
from $H^\kappa$ to itself by the same argument with $U(t,s)$
replaced by $\widetilde{U}(r,s)$.  The case of $|t|<T, s < -T$ is
similar.

\section{Asymptotic completeness in Sobolev spaces}

Recall that we are considering \eqref{eq:2pot}. $V_1$ is
stationary (we denote its velocity as $\vec e_0=0$) and $V_2$ is
moving with velocity $\vec e_1$. There are two approaches to prove
Theorem~\ref{thm:main4}. Graf (\cite{Gr}) proved the asymptotic
completeness for the charge transfer model in the $L^2$
 sense by proving a RAGE theorem. Our first option to prove
 Theorem~\ref{thm:main4} is to generalize
Graf's idea. We find that this approach works, provided the fact
that each individual subsystem (i.e. $p^2 + V_l$) is
asymptotically complete in the $H^\kappa$ sense. However the only
direct way to prove this fact, as we know, is by the dispersive
estimate. The good point of this approach is that it requires less
restrictive condition on the potentials and the spectrum of the
individual subsystem, given that nontrivial fact. Our second
option to prove Theorem~\ref{thm:main4} is to apply the dispersive
estimate (Theorem~\ref{thm:main2}) directly. To illustrate both of
these ideas, the following proof is somehow a combination of these
two options. Specifically, we follow \cite{Gr} to prove the
existence of the wave operators and then apply
Theorem~\ref{thm:main2} to prove Theorem~\ref{thm:main4}.

\subsection{Existence of wave operators}
 The well-known wave operators are defined as
following:

$$
\Omega^-_{0}(s) = \underset{ t\to +\infty}{s-\lim}\, U(s,t)
e^{-i(t-s)H_0},$$

$$
\Omega^-_{1}(s) = \underset{ t\to +\infty}{s-\lim}\, U(s,t)
e^{-i(t-s)H_{1}} P_b(H_1),$$

$$
\Omega^-_2(s) = \underset {t\to +\infty }{s-\lim }\, U(s,t)
\calg_{-\vec e_{1}}(t) e^{-i(t-s)H_2} P_b (H_2)\calg_{\vec
e_{1}}(s).$$

\begin{theorem}
Under the assumption of Theorem~\ref{thm:main3}, the above wave
operators exist in the space $H^\kappa$. More precisely, for
$l=0,1,2$ and $\forall \psi_0 \in H^\kappa$, the limits converges
in the $H^\kappa$ sense and $\Omega^-_{l}(s)\psi_0$ lies in
$H^\kappa(\R^n)$. \label{thm:wave}
\end{theorem}

\begin{remark} The above theorem can be proved by Cook's method
together with Theorem~\ref{thm:main2} and Theorem~\ref{thm:main3}
if we are willing to impose more regularity on the potentials and
the spectrum condition. The following proof originated in
\cite{Gr}, which we believe, requires the least conditions on the
system.
\end{remark}

We present some preliminary facts before we proceed:

\begin{lemma} \label{lemma:wave1}
Let $g \in C^\infty_0(\R^n)$ and $\nu > 0$. Suppose

\begin{enumerate}

\item  $ g(p)=0 $ for $|p|\geq \nu$  and fix $\alpha > 1$. Then
for $R>0, t>0$ and any $N>0,$
$$
\| F(|x|> \alpha(R+\nu t))e^{-i\frac{p^2}2
t}g(p)F(|x|<R)\psi\|_{H^\kappa} \leq C_{N,\kappa}(R+\nu t)^{-N} \|
\psi\|_{L^2}.
$$

\item  $ g(p)=0 $ for $|p|\leq \nu$ and $\nu_0 >0, \, 0<\alpha
<1$. Then for $t>0$ and any $N>0,$
$$
\| F(|x|< \alpha(\nu-\nu_0) t)e^{-i\frac{p^2}2 t}g(p)F(|x|<\nu_0
t) \psi \|_{H^\kappa} \leq C_{N,\kappa}t^{-N} \| \psi\|_{L^2}.
$$
\end{enumerate}

\end{lemma}

These estimates are fairly common for $\kappa = 0$ and may be
proved by the stationary phase methods (e.g. \cite{En}, Lemma
(6.3)). For the case $\kappa \geq 1$, it follows from a commutator
argument and the fact that the derivative on the left-hand side
can be absorbed into $g(p)$ because $g \in C^\infty_0(\R^n)$. The
next lemma represents to some extent the counterpart of
Lemma~\ref{lemma:wave1} for $H_l = H_0 + V_l$.

\begin{lemma} \label{lemma:wave2}
Let $g \in C^\infty_0(\R)$ and $v > 0$. Suppose $g(e)=0$ for
$e\geq v^2/2$ and fix $\alpha > 1$. Then for $l=1,2$, $R>0 $ and
$t\geq 0$, we have

\be \| F(|x|> \alpha(R+v t))e^{-iH_l
t}g(H_l)F(|x|<R)\psi(x)\|_{H^\kappa} \leq C_{N,\kappa}(R+v
t)^{-\epsilon} \| \psi\|_{L^2} . \label{eq:wave2} \ee
\end{lemma}

When $\kappa =0$, the lemma is just Lemma 4.2 of \cite{Gr}. For
$\kappa \geq 1$, the left-hand side of \eqref{eq:wave2} is
dominated up to a constant by
$$
\| (H_l + M)^{\frac {\kappa}2} e^{-iH_l
t}g(H_l)F(|x|<R)\psi(x)\|_{L^2 (|x|> \alpha(R+v t))}.
$$

\noindent where M is chosen so large that $H_l + M$ is a positive
operator. If we define $\widetilde{g}(H_l)= (H_l+M )^{\frac
{\kappa}2} g(H_l) $, then $\widetilde{g} \in C^\infty_0(\R)$. The
above is of the form of $\kappa =0$ and the lemma follows from the
case $\kappa = 0$.

\begin{lemma} \label{lemma:wave3}
\begin{enumerate}
\item  Let $0<v_0<v$ and $g \in C^\infty_0(\R^n)$ with $g(p)=0$
for $\{ |p|< v \} \bigcup \{ |p- \vec e_1|< v \} $. Then for any $
s \in \R $,

$$ \underset{ t_1\to +\infty}{\lim} \underset{ t_2>
t_1}{\sup}\,\|(U(t_2,t_1)-e^{-i H_0(t_2-t_1)})e^{-i H_0(t_1-s)}
g(p) \prod_{l=0}^1 F(|x- \vec e_l s|<v_0(t_1-s)) \|_{L^2\to
H^\kappa} =0.
$$

\item  Let $ v_0,v>0$ with $v_0+v < |\vec e_1|$ and $g \in
C^\infty_0(\R)$ with $g(p)=0$ for $p > v^2/2$. Then

$$ \underset{ t_1\to +\infty}{\lim} \underset{ t_2>
t_1}{\sup}\,\|(U(t_2,t_1)-e^{-i H_1(t_2-t_1)})g(H_1) F(|x|<v_0
t_1) \|_{L^2\to H^\kappa} =0 . $$

\end{enumerate}
\end{lemma}

For $\kappa=0$, the lemma was proved in \cite{Gr}. We will follow
the approach there to prove the case $\kappa >0$.

\begin{proof} Part (1):
Take $\alpha<\alpha_1<1$ and let $f\in C^\infty_0(R^n) $ with
$f(y)=0$ if $|y-\vec e_l| > \alpha(v-v_0)$ for both $l=0,1$. Since
$\alpha t < \alpha_1 (t-s)$, we have $ |f(x/t)| \leq |f(x/t)|
\sum_{l=0}^1 F(|x-\vec e_l t|< \alpha_1 (v-v_0)(t-s)) $ for $t$
large enough.

\begin{align} &
\|f(\frac xt)e^{-i H_0(t-s)} g(p) \prod_{l=0}^1 F(|x-\vec e_l
s|<v_0(t-s))\|_{L^2 \to H^\kappa} \nn \\
& \lesssim \sum_{|\beta|<\kappa}\sum_{l=0}^1 \|\partial^\beta
f\|_\infty  \|F(|x-\vec e_l t|< \alpha_1 (v-v_0)(t-s))\nn \\
 & \quad \qquad \qquad e^{-i H_0(t-s)}
g^\beta(p) F(|x-\vec e_l s|<v_0(t-s))\|_{L^2 \to L^2 } \label{eq:wave3} \\
& \leq C \sum_{|\beta|<\kappa}\sum_{l=0}^1 \|F(|x|< \alpha_1
(v-v_0)(t-s))e^{-i H_0(t-s)} g^\beta(p+\vec e_l) F(|x|<v_0(t-s))\| \nn \\
& \leq C (t-s)^{-N} \nn
\end{align}

\noindent where $g^\beta(p)= \sum_{|\beta + \gamma|=\kappa}
p^{\gamma} g(p) $. The above inequality follows by commuting the
derivative through $f(x/t)$, by applying Galilean transform to the
second expression, and by Lemma~\ref{lemma:wave1}. By
\eqref{eq:wave3} and Theorem~\ref{thm:main3}, it suffices to show

\begin{align}   \underset{ t_2> t_1}{\sup} \, \|(U(t_2,t_1) &
(1-f(x/t_1))-(1-f(x/t_2)) e^{-i H_0(t_2-t_1)})e^{-i H_0(t_1-s)}
g(p) \cdot \nn
\\  & \cdot \prod_{l=0}^1 F(|x-\vec e_l
s|<v_0(t_1-s)) \|_{L^2\to H^\kappa} \to 0. \label{eq:wave4}
\end{align}

Substituting $$(U(t_2,t_1)(1-f(x/t_1))-(1-f(x/t_2))e^{-i
H_0(t_2-t_1)})= \int_{t_1}^{t_2} \frac d{dt}(
U(t_2,t)(1-f(x/t))e^{-i H_0(t-t_1)})dt $$  into \eqref{eq:wave4},
it follows from Theorem~\ref{thm:main3} that the left-hand side of
\eqref{eq:wave4} is dominated by
$$
\int_{t_1}^{+\infty} dt \|[i H(t)(1-f(x/t))-i(1-f(x/t))H_0 - \frac
{\partial}{\partial t}f(x/t)]e^{-i H_0(t-s)} g(p) \prod_{l=0}^1
F(|x-\vec e_l s|<v_0(t_1-s))\|_{L^2\to H^\kappa}.
$$
\noindent The expression within the square brackets consists of
(1)-(3)  which are estimated as follows:
\begin{enumerate}
\item Suppose $t$ is sufficiently large, then $V_l(t)(1-f(x/t))=0
$, because $V_l$ is compactly supported,
 where we take $f(y)=1$ for $|y-\vec e_l|< \alpha (v-v_0)/2$.

\item $H_0 f(x/t)-f(x/t) H_0 = -\frac 12 t^{-2} (\triangle f)(x/t)
-i t ^{-1}(\nabla f)(x/t)p $ and

\item  $\frac {\partial}{\partial t} f(x/t)=-t^{-1}(x/t)(\nabla f)
(x/t)$
\end{enumerate}
\noindent are treated using \eqref{eq:wave3}.

Part (2): Choose $\alpha >1 $ and $v_1$ with $\alpha (v+v_0)<v_1 <
|e_1|$ and let $f \in C^\infty_0(\R^n)$ with $f(y)=1$ for $|y|<
\alpha (v+v_0)$ and $f(y)=0$  for $|y|> v_1$. We first claim that

\be \underset{t_1 \to +\infty }{\lim} \underset{ t> t_1}{\sup}
\|(1-f(x/t)) e^{-i H_1(t-t_1)})g(H_1) F(|x|<v_0 t_1) \|_{L^2\to
H^\kappa}=0 \ee

Since $1-f(x/t)$ is supported in  $|x|> \alpha (v+v_0)t > \alpha
(v_0 t_1 +v(t-t_1))$, it follows from Lemma~\ref{lemma:wave2} that
\be \label{eq:wave5} \| F(|x|> \alpha(v_0 t_1+v (t-t_1))e^{-iH_l
(t-t_1)}  g(H_l)F(|x|<v_0 t_1 )\|_{L^2 \to H^\kappa} \leq
C_{N,\kappa}(v_0 t_1+v (t-t_1))^{-\epsilon}. \ee

Now by Theorem~\ref{thm:main3} and
$$(U(t_2,t_1)f(x/t_1)-f(x/t_2)e^{-i H_1(t_2-t_1)})=
\int_{t_1}^{t^2} \frac d{dt}(U(t_2,t)f(x/t)e^{-i H_1 (t-t_1)}
)dt,$$ it suffices to estimate

\begin{align*}  & \underset{ t_2>
t_1}{\sup}\,\|(U(t_2,t_1)f(x/t_1)-f(x/t_2)e^{-i
H_1(t_2-t_1)})g(H_1) F(|x|<v_0 t_1) \|_{L^2\to H^\kappa} \\
& \leq \int_{t_1}^{+\infty}dt \, \|[iH(t)f(x/t)- i f(x/t)H_1
+\frac {\partial}{\partial t }f(x/t)]e^{-i H_1(t-t_1)}g(H_1)
F(|x|<v_0 t_1)\|_{L^2\to H^\kappa},
\end{align*}

As in Part (1), a discussion of terms (a)-(d) in the square
brackets now follows:

(a)  $V_1(x)f(x/t)-f(x/t)V_1(x)=0$

(b)  $V_2(x-e_1 t) f(x/t)=0$ if $t$ is large enough because $V_2$
is compactly supported and $f(y)=0$ for $|y|>v_1$ and $|e_1|>
v_1$.

(c)  $[H_0, f(x/t)]=\frac 12 t^{-2}\triangle f (x/t)- \frac {ip}t
\nabla f(x/t)$. Since $V_1 \in C^\kappa_0 $, we can take $M$ large
enough, so that the corresponding term can be dominated by

$$ \|(M+H_1)^{\frac \kappa 2} (\frac 12 t^{-2}\triangle f
(x/t)- \frac {ip}t \nabla f(x/t)) e^{-i H_1(t-t_1)}g(H_1)
F(|x|<v_0 t_1)\|_{L^2 \to L^2}. $$

Commute $(M+H_1)^{\frac \kappa 2}$ through $ (\frac 12
t^{-2}\triangle f (x/t)+ \frac 1t \nabla f(x/t) \nabla )$ and the
commutators generated will decay at least as fast as $t^{-2}$,
hence they are time-integrable. Note $\|(p^2 +1)^\sigma g(H_1)
\|_{L^2 \to L^2} < C_\sigma $. The only term that does not decay
as fast as $t^{-2}$ is

$$ \|(M+H_1)^{-1}(\frac {ip}t \nabla f(x/t)) e^{-i H_1(t-t_1)}(M+H_1)^{\frac \kappa 2 +1}g(H_1)
F(|x|<v_0 t_1)\|_{L^2 \to L^2} ,$$

\noindent which is integrable, due to the fact that $(M+H_1)^{-1}p
$ is a bounded operator from $L^2$ to $L^2$ and due to
\eqref{eq:wave5} (with $g(H_1)$ replaced by $(M+H_1)^{\frac \kappa
2 +1}g(H_1)$), and due to the support property of $ \nabla
f(x/t)$.

(d) $\frac {\partial}{\partial t}f(x/t) = -\frac xt f(x/t)t^{-1},
$ which can be treated as part (c), using \eqref{eq:wave5} with
$f(x)$ replaced by $x f(x)$.
\end{proof}

\begin{proof}[proof of Theorem~\ref{thm:wave}]
Since $U(s,t) e^{-i(t-s)H_0}$ and $U(s,t) e^{-i(t-s)H_1}$ are
uniformly bounded operators from $H^\kappa$ to $H^\kappa$, it
suffices to prove the existence of the strong limits
$\Omega^-_{0}(s)$ and $ \Omega^-_{1}(s)$ on a dense set $D$:

$$D=\{ g(p)f(x)\psi : g\in C_0^\infty(\R^n \backslash \{0,e_1 \} ),
 f \in C_0^\infty(\R^n), \psi \in L^2(\R^n)\} .$$

$g(p)$ satisfies the hypothesis of Lemma~\ref{lemma:wave3} Part
(1), with a suitable $v >0$. Take $0<v_0<v$ and note that

$$\prod_{l=1}^2 F(|x-\vec e_l|<v_0(t_1-s)) f(x)=f(x) $$
\noindent for $t_1$ big enough. For $t_2> t_1$ , it follows from
Theorem~\ref{thm:main3} that

\begin{align*} & \|(U(s,t_1)e^{-i H_0(t_1-s)} - U(s,t_2) e^{-i H_0(t_2-s)}) g(p)
f(x) \psi \|_{H^\kappa} \\
\lesssim & \|(U(t_2,t_1)-e^{-i H_0(t_2-t_1)})e^{-i H_0(t_1-s)}
g(p) \prod_{l=0}^1 F(|x-\vec e_l s|<v_0(t_1-s))\|_{L^2\to
H^\kappa}\|f(x)\psi\|_{L^2}.
\end{align*}

\noindent Hence Lemma~\ref{lemma:wave3} implies that $U(s,t )
e^{-i H_0(t -s)} g(p) f(x) \psi $ is Cauchy sequence in $H^\kappa
(\R^n)$ as $t \to +\infty$, which is equivalent to the existence
of $\Omega^-_{0}(s) $ .

We will only show the existence of $ \Omega^-_{1}(s)$. The
existence of $ \Omega^-_{2}(s)$ follows from the same argument up
to a Galilean transform (\cite{Gr}). Since the eigenfunctions of
$H_1$ span the range of $P_b(H_1)$, it suffices to prove
convergence on the eigenfunctions $ \psi: \,\, H_1 \psi = E \psi
$.  Due to our assumptions on the potentials, the positive
eigenvalues are excluded. Thus for any $v>0$, we can find a
suitable $g$ as in Lemma~\ref{lemma:wave3} Part (2) with $g(H_1)
P(H_1) = P(H_1)$. More precisely, we take $v, v_0 >0$ with $v+v_0
< |e_1|$. For $t_2 \geq t_1$,
\begin{align*} &
\|(U(s,t_1)e^{-i H_1(t_1-s)}P(H_1) - U(s,t_2) e^{-i H_1(t_2-s)})
 \psi \|_{H^\kappa} \\
=&  \|U(s,t_2)(U(t_2,t_1)- e^{-i H_1(t_2-t_1)} )  e^{-i
H_1(t_1-s)} g(H_1)(F(|x|<v_0 t_1)+F(|x|>v_0 t_1)) \psi \|_{H^\kappa} \\
\lesssim &   \|(U(t_2,t_1)- e^{-i H_1(t_2-t_1)}) e^{-i H_1(t_1-s)}
g(H_1)F(|x|<v_0 t_1)\|_{L^2 \to H^\kappa} \|\psi\|_{L^2} + \|
F(|x|>v_0 t_1)\psi \|_{H^\kappa},
\end{align*}
\noindent since $ U(s,t),H_1(s) $ and $g(H_1)$ are bounded
operators on $H^\kappa (\R^n)$ with a uniform bound.

Lemma~\ref{lemma:wave3} part (2) and the fact that $\| F(|x|>v_0
t_1)\psi \|_{H^\kappa} \to 0$ when $t_1 \to +\infty$ imply that
$U(s,t )e^{-i H_1(t-s)}P(H_1) \psi$ is a Cauchy sequence in
$H^\kappa$.
\end{proof}

\subsection{Asymptotic completeness}
In this section we will apply Theorem~\ref{thm:main2} and
\ref{thm:main3} to prove Theorem~\ref{thm:main4}. For the case
$\kappa =0$, we refer the reader to \cite{RSS1}.

\textit{Proof of Theorem \ref{thm:main4}:} First let us assume
that $\psi_0 \in W^{\kappa,2} \cap W^{\kappa,p'}$ for some $1<p'<
\frac {2n}{2+n}$. Decompose
$$
\psi(t):=U(t) \psi_0 = P_b(H_1) U(t) \psi_0 +  P_b (H_2,t) U(t)
\psi_0 + R(t).
$$
By construction, we clearly have
\begin{align}
& P_b(H_2,t) U(t) \psi_0 + R(t) \in \Ran(P_c (H_1)),\label{eq:dec2}\\
&P_b(H_1) U(t) \psi_0 + R(t) \in \Ran(P_c(H_2,t)).\nn
\end{align}
We further write
$$
P_b(H_1) U(t) \psi_0 = \sum^m_{r =1} e^{-i\lambda_r t} a_r (t) u_r
(x)
$$
for some choice of unknown functions $a_{r}(t)$. Due to the
smoothness of the potentials, $u_r$ belongs to $H^\kappa(\R^n)$.
It follows from \eqref{eq:dec2} that, similar to \eqref{eq:ODE},
$$
\dot a_{r}+ i\,\langle V_2 (\cdot - t\vec e_{1})\psi(t),
u_r\rangle = 0 \text{\ \ for all\ \ }1\le r\le m.
$$
The exponential localization of $u_r$ implies that $ |\langle V_2
(\cdot - t\vec e_1) \psi(t), u_{r} \rangle | \les e^{-\alpha t}$.
Therefore, $a_r (t)$ has a limit, writing $ \lim_{t\to +\infty}
a_r(t) =A_{r},$ and
\begin{equation}
\label{eq:tired} \Big\| P_b (H_1) U(t) \psi_0 - \sum^m_{r=1} A_r
e^{-i\lambda_r t}u_r \Big\|_{H^\kappa}\to 0, \quad t\to +\infty.
\end{equation}
We next define the functions $v_r = \underset{ t\to
+\infty}{\lim}\,  U(t)^{-1}  e^{-i\lambda_r t} u_r.$ The existence
of $v_r$ and $v_r \in H^\kappa $ is guaranteed by
Theorem~\ref{thm:wave}. By Theorem~\ref{thm:main3}, we have
\begin{equation}
\label{eq:1chan} \Big\|U(t) \big (\sum^m_{r=1} A_{r} v_r\big ) -
\sum^m_{r=1} A_r e^{-i\lambda_r t}u_r \Big\|_{H^\kappa}\to 0,
\quad t \to + \infty.
\end{equation}
We then infer from \eqref{eq:tired} that
\begin{equation}
\label{eq:pbh1} \Big\|U(t) \big (\sum^m_{r=1} A_{r} v_r\big ) -
P_b (H_1) U(t) \psi_0 \Big\|_{H^\kappa}\to 0, \quad t \to +
\infty.
\end{equation}
The above arguments apply to $P_b(H_2,t) U(t) \psi_0$ in a similar
fashion. More precisely, we write
$$
U(t) \psi_0 = P_b (H_2,t) U(t) \psi_0 + \Gamma(t) =\calg_{-\vec
e_{1}}(t)P_b (H_2) \calg_{\vec e_{1}}(t) U(t) \psi_0 + \Gamma(t).
$$
Therefore,
\begin{equation}
\label{eq:2dec}
 \calg_{\vec e_{1}}(t)U(t) \psi_0 = P_b (H_2) \calg_{\vec e_{1}}(t)
 U(t) \psi_0 + \calg_{\vec e_{1}}(t)\Gamma(t)
\end{equation}
Recall that the function $ \tilde \psi(t) = \calg_{\vec e_{1}}(t)
U(t) \psi_0$ is a solution of the problem
\begin{equation}
\label{eq:anduh} \frac{1}{i} \partial_t \tilde \psi -
\Laplace\tilde \psi + V_2(x) \tilde\psi + V_1(x + t\vec e_1)
\tilde\psi = 0,\qquad \tilde\psi|_{t=0} = \calg_{\vec
e_{1}}(0)\psi_{0} .
\end{equation}
According to \eqref{eq:2dec}, $\tilde\psi(t) = P_b (H_2) \tilde
\psi(t) +  \Gamma_1 (t)$, where $\Gamma_1 (t) = \calg_{\vec
e_{1}}(0) \Gamma(t)$. In particular,
$$
\Gamma _1(t) \in Ran (P_c (H_2)).
$$
Decompose
$$
P_b(H_2) \tilde\psi(t) = \sum^\ell_{s=1} b_s (t) e^{-i \mu_{s}t}
w_s
$$
for some choice of unknown functions $b_{s}(t)$. Again due to the
smoothness of the potentials, $w_s \in H^\kappa(\R^n)$. After
substituting the decomposition in \eqref{eq:anduh} we obtain the
equations
$$
\dot b_s (t) + i \,\langle V_1 (\cdot + t\vec e_1) \tilde\psi ,
w_s\rangle = 0 \text{\ \ for all\  \ }1\le s\le\ell.
$$
Using exponential localization of $w_s$ we conclude the existence
of the limit $b_s(t) \to B_{s}$ as $t \to +\infty$. Thus $\| P_b
(H_2) \tilde\psi(t) - \sum ^\ell_{s=1} B_{s} e^{-i\mu_s t} w_s
\|_{H^\kappa} \to 0,  \quad t\to \infty $. Equivalently, after
applying $\calg_{-\vec e_{1}}(t)$, we have
\begin{equation}
\label{eq:asdec} \Big\|  P_b (H_2,t) U(t) \psi_0 - \sum^\ell_{s=1}
B_{s} e^{-i\mu_j t} \calg_{-\vec e_{1}}(t) w_s \Big\|_{H^\kappa}
\to 0.
\end{equation}
Now Theorem~\ref{thm:wave} allows us to define
$$\omega_s :=
\Omega^-_2 w_s = \underset {t\to +\infty }{s-\lim }\, U(t)^{-1}
\calg_{-\vec e_{1}}(t) e^{-itH_2} P_b (H_2)w_s \in H^\kappa .$$

Moreover,
\begin{equation}
\label{eq:2chan} \Big\|U (t) \big (\sum_{s=1}^{\ell}
B_{s}\omega_s\big ) - \sum_{s=1}^{\ell } B_{s} e^{-i\mu_s t}
\calg_{-\vec e_{1}} (t) w_s \Big\|_{H^\kappa} \to 0,  \quad t\to +
\infty.
\end{equation}
It then follows from \eqref{eq:asdec} that
\begin{equation}
\label{eq:pbh2} \|  P_b (H_2,t) U(t) \psi_0 - U(t) \big
(\sum^\ell_{s=1} B_s \omega_s\big ) \|_{H^\kappa} \to 0, \quad t
\to + \infty.
\end{equation}
We now define the function
\begin{equation}
\label{eq:free} \phi := \psi_0 - \sum^m_{r=1} A_r v_r -
\sum^\ell_{s=1} B_s \omega_s,
\end{equation}
which will lead to the initial data $\phi_{0}$ for the free
channel. We have that
$$
P_b(H_1) U(t) \phi = P_b(H_1) U(t) \psi_0 - P_b (H_1) U(t) \big
(\sum^m_{r=1} A_{r} v_r\big ) - P_b (H_1) U(t) \big
(\sum^\ell_{s=1} B_s \omega_s\big ).
$$
It follows from \eqref{eq:pbh1} and the identity $P^{2}_{b}(H_{1})
= P_{b} (H_{1}) $  that
\begin{equation}
\label{eq:again1} \Big\| P_b(H_1) U(t) \psi_0 - P_b (H_1) U(t)
\big (\sum^m_{r=1} A_{r} v_r\big ) \Big\|_{H^\kappa} \to 0 \qquad
\text{as}\,\, t\to +\infty.
\end{equation}
Furthermore,
\begin{equation}
\label{eq:again2} P_b(H_1) \sum^\ell_{s=1} B_s e^{-i\mu_s t}
\calg_{-\vec e_{1}}(t) w_j =\sum^m_{r=1} \sum^\ell_{s=1} B_s
e^{-i\mu_s t} \langle
 \calg_{-\vec e_{1}}(t) w_j , u_r \rangle \, u_{r}\to 0
\end{equation}
in the $H^\kappa$ sense as $t\to +\infty $, due to the exponential
localization of the eigenfunctions $u_{r}$. We infer
from~\eqref{eq:again1}, \eqref{eq:2chan}, and~\eqref{eq:again2}
that $ \|P_b (H_1) U(t) \phi \|_{H^\kappa} \to 0$. Similarly,
$\|P_b (H_2,t) U(t) \phi\|_{H^\kappa} \to 0$. Thus, $U(t) \phi$ is
asymptotically orthogonal to the bound states of $H_1$ and $H_2$.
$V_j \in C_0^{n+2\kappa +2}$ implies that $(1+|\xi|)^{\kappa
+1+\frac n2} \widehat{V}_j (\xi) \in L^2(\R^n) $. So $
(1+|\xi|)^{\kappa} \widehat{V}_j (\xi) \in L^1(\R^n) $. Therefore,
according to Theorem~\ref{thm:main2}, $U(t) \phi$ satisfies the
estimate

\begin{equation}  \label{eq:wavedec}
\| U(t) \phi \|_{W^{\kappa,p}} \lesssim |t|^{-n (\frac 12-\frac
1p)} \|\phi\|_{W^{\kappa,p'}}
\end{equation}

\noindent where $\frac {2n}{n-2} <p<+\infty $. In order to be able
to apply the estimate \eqref{eq:wavedec}, one needs to verify that
$\phi\in W^{\kappa,p'}$. By assumption, $\psi_{0}\in
W^{\kappa,p'}$. Thus it remains to check $v_{r}\in
W^{\kappa,p'},\, r=1,..,m$ and $\omega_{s}\in W^{\kappa,p'},\,
s=1,..,\ell$, which is guaranteed by Lemma~\ref{lemma:woper}
below. Assuming this lemma for the moment, we now consider the
expression
$$
e^{-it\Laplace}U(t)\phi = \phi -i \int^t_0 e^{-is\Laplace}
\left(V_1(x) + V_2 (x- s \vec e_1)\right) U(s) \phi\, ds.
$$
Writing $\frac {2p}{p-2}=r$, we have the following estimate:
\begin{align*} &
\int^{+\infty}_t \|e^{-is\Laplace} \left( V_1(x) + V_2(x - s\vec
e_1)\right)U(s) \phi\|_{H^\kappa} ds \\
& \lesssim (\|V_1\|_{W^{\kappa,r}} + \|V_2\|_{W^{\kappa,r}})\int^{+\infty}_t
\| U(s) \phi \|_{W^{\kappa,p}} ds\\
&\les \int^{+\infty}_t |s|^{-n (\frac 12-\frac 1p)}
\|\phi\|_{W^{\kappa,p'}} (\|V_1\|_{W^{\kappa,r}} +
\|V_2\|_{W^{\kappa,r}})  \to 0, \quad \text{as}\,\,\,\, t\to
+\infty.
\end{align*}
\noindent  Here we note that $-n (\frac 12-\frac 1p) <-1 $. This
allows us to show the existence of the limit
$$
\phi_{0}:=\underset{ t\to\infty}{\lim} e^{it\Laplace} U(t) \phi
\in H^\kappa.
$$
It follows that
\begin{equation}
\label{eq:freech} \|U(t)\phi - e^{-it\Laplace} \phi_0\|_{H^\kappa}
\to 0, \quad  t\to + \infty.
\end{equation}
Combining \eqref{eq:1chan}, \eqref{eq:2chan}, \eqref{eq:free}, and
\eqref{eq:freech} we infer that
$$
\Big\|U(t)\psi_0 - \sum^m_{r=1} A_r e^{-i\lambda_r t} u_r -
\sum^\ell_{s=1} B_{s} e^{-i\mu_s t}\calg_{-\vec e_{1}}(t) w_s
 - e^{-it\Laplace} \phi_0 \Big\|_{H^\kappa} \to 0,\quad  \text{
as }\,\,\, t\to + \infty,
$$
as claimed. Because $W^{\kappa,2} \cap W^{\kappa,p'}$ is dense in
$W^{\kappa,2}$, for any $\psi_0 \in W^{\kappa,2}$, there is a
sequence $\psi_l \in W^{\kappa,2} \cap W^{\kappa,p'} $ converging
to $\psi_0$ in the $ W^{\kappa,2}$ norm. Then for each $\psi_l$,
we have the following decomposition:

\begin{equation}
\nn U(t) \psi_l = \sum^m_{r=1} A_r^l e^{-i\lambda_rt} u_r +
\sum^\ell_{k=1}B_k^l e^{-i\mu_k t} \calg_{-\vec e_{1}} (t) w_k +
e^{-it\Laplace} \phi_l + \mathcal {R}_l(t),
\end{equation}

It follows from Theorem~\ref{thm:main3} that $\psi_l
=\sum^m_{r=1}A_r^l \Omega^-_{1} u_r + \sum^\ell_{k=1}B_k^l
\Omega^-_{2} w_k  + \Omega^-_{0} \phi_l.$

Since the ranges of $\Omega^-_{0,1,2}$ are orthogonal to each
other in $L^2(\R^n)$ (\cite{Gr}), the fact that $\psi_l$ converges
as $l \to +\infty$, implies that each component in the above
equation converges. Hence, $\lim_{l \to +\infty} A_r^l= A_r^0 $,
$\lim_{l \to +\infty} B_k^l= B_k^0 $. These imply that $
\Omega^-_{0} \phi_l$ converges in $H^\kappa$, since all other
terms in the above identity converges in $H^\kappa$. Write
$\lim_{l \to +\infty}\Omega^-_{0} \phi_l=f_0 \in H^\kappa$.

By the asymptotic completeness theorem for $L^2$ (\cite{Gr}),
there are $\phi_0 \in L^2 $ such that the following holds:

$$\psi_0 =\sum^m_{r=1}A_r^0 \Omega^-_{1} u_r + \sum^\ell_{k=1}B_k^0
\Omega^-_{2} w_k  + \Omega^-_{0} \phi_0$$ in the $L^2$ sense. This
implies that $f_0 =\Omega^-_{0} \phi_0 $.

Then by the definition of the wave operator, $ U(0,t) e^{-itH_0}
\phi_0 - f_0 \to 0$ as $t \to +\infty$ in $L^2(\R^n)$. Since
$U(t,0)$ and $e^{itH_0} $ are uniformly bounded operators on
$H^\kappa(\R^n)$ and $L^2(\R^n)$, we see that $\phi_0 = \underset
{t \to +\infty}{\lim} e^{itH_0} U(t,0) f_0 $ in $L^2(\R^n)$. We
claim that this implies $\phi_0 \in H^\kappa(\R^n)$. It suffices
to prove the following:

Assume $ g_n$ is a sequence in $H^\kappa(\R^n)$ and $
\|g_n\|_{H^\kappa} <1$. Moreover, $g_n$ converges to $g$ in the
$L^2$ norm. Then $g $ lies in $H^\kappa(\R^n)$.

To see this, note that on Fourier side, $H^\kappa(\R^n)$ is just a
weighted $L^2(\R^n)$ space. More precisely, $ \|\hat {g_n}- \hat
{g}\|_{L^2(\R^n)} \to 0 $ implies that for the ball $B_R$ with
radius $R$, centered at the origin,
$$ \| (1+|\xi|^2)^{\frac {\kappa}2} (\hat {g_n}(\xi)- \hat
{g}(\xi))\|_{L^2(B_R)} \to 0 \quad \text {as} \quad n \to +\infty
.$$ This implies that $\| (1+|\xi|^2)^{\frac {\kappa}2}  \hat
{g}(\xi)\|_{L^2(B_R)}$ is uniformly bounded by $\sup
\|g_n\|_{H^\kappa}  \leq 1 $. Let $R \to +\infty$, we see that
$\|g\|_{H^\kappa} \leq 1 $.

Now it is clear that the following decomposition holds in the
space $H^\kappa$ for any $\psi_0 \in H^\kappa$:
$$\psi_0 =\sum^m_{r=1}A_r^0 \Omega^-_{1} u_r + \sum^\ell_{k=1}B_k^0
\Omega^-_{2} w_k  + \Omega^-_{0} \phi_0.$$

To complete the proof of Theorem~\ref{thm:main4}, it remains to
prove the following lemma:

\begin{lemma}
Assume that the potentials $V_{1}(x), V_{2} \in C_0^{
n+2\kappa+2}(\R^n)$. Let $U(t)$ be the evolution operator of
\eqref{eq:2pot} and $ \Omega_{1,2}^{-}$ the wave operators
corresponding to $U(t)$, as defined at the beginning of this
section.  Then for $ \forall f \in L^2(\R^n)$, $\Omega_{1,2}^{-}f$
lies in $W^{\kappa,p'}$, where $ 1<p'<\frac{2n}{n+2}$.
\label{lemma:woper}
\end{lemma}

\begin{proof} The proof is essentially contained in
\cite{RSS1} Section 4. For the reader's convenience, we present
the details here. Without loss of generality we only consider the
wave operator $ \Omega_{1}^{-}$. For an arbitrary $L^{2}$ function
$f$
$$
\Omega_{1}^{-} f = \sum_{r=1}^{m} f_{r}\underset{t\to
+\infty}{\lim} U(t)^{-1} e^{-it H_{1}} u_{r},
$$
where $P_{b}(H_{1}) f =\sum_{r=1}^{m} f_{r} u_{r}$ for some
constants $f_{r}$. It follows from Duhamel's formula that
\begin{align}
U(t)^{-1} e^{-itH_{1}} u_{r} = &u_{r} + i \int_{0}^{t} U(s)^{-1}
V_{2}(\cdot - s\vec e_{1}) e^{-isH_{1}} u_{r}\, ds \nn\\ = &u_{r}
+ i \int_{0}^{t} U(s)^{-1} V_{2}(\cdot - s\vec e_{1})
e^{-i\lambda_{r} s} u_{r}\, ds, \label{eq:duhamel}
\end{align}
since $u_{r}$ is an eigenfunction of $H_{1}$ corresponding to an
eigenvalue $\lambda_{r}$. The function $u_{r}$ is exponentially
localized in $L^{2}$ together with its $n+2$ derivatives \footnote
{The localization of higher derivatives of $u_{r}$ follows from
the localization of $u_{r}$ stated in \eqref{eq:local} and the
equation $-\Laplace u_{r} + V_{1}(x) u_{r} = \lambda_{r} u_{r}$
with potential $V_{1}(x)$ which is bounded together with all its
derivatives of order $\le (n+2)$.}
$$
\sum_{0\le |\gamma|\le n+2} \int_{\R^{n}}e^{2\alpha |x|}
|\partial_{x}^{\gamma}u_{r}(x)|^{2}\, dx \le C
$$
for some positive constant $\alpha$ appearing in \eqref{eq:local}.
This implies that the function
$$
G_{r}(s,x): = e^{-i\lambda_{r} s} V_{2}(x-s\vec e_{1}) u_{r} (x)
$$
has the property that for any $k\ge 0$ and multi-index $\gamma$,
$0\le |\gamma|\le n+2$
$$
\|\la x\ra^{k}\partial^{\gamma}_{x}
G_{r}(s,\cdot)\|_{L^{2}_{x}}\le c(r,|\gamma |, k) \la
s\ra^{-3j_0-2-\kappa}.
$$
By H\"older's inequality,we have, writing $q= \frac {2p'}{2-p'}$,
\be
 \|\partial^{\gamma}_{x}U(-s)G(s,x)\|_{L^{p'}} \leq
 \| \la x\ra^{-j_0} \|_{L^q} \|\la x\ra^{j_0}
 \partial^{\gamma}_{x}U(-s)G(s,x)\|_{L^2}.
\ee

Take $j_0=[\frac {2-p'}{2p'}n ] +1 > \frac nq$, then $\| \la
x\ra^{-j_0} \|_{L^q} < +\infty$.

To prove the desired conclusion, it would then suffice to show
that for any $|\gamma|=\kappa $, there exists a positive constant
$k$ such that for any function $g(x)$
\begin{equation}
\label{eq:invariance} \|\la x\ra^j \partial^\gamma   U(t) g
\|_{L^{2}} \les \la t\ra^{3j_0 +\kappa}\sum_{|\beta|\le j_0
+\kappa} \|\la x\ra ^{k}
\partial_{x}^{\beta} g\|_{L^{2}},\qquad \forall t\ge 0.
\end{equation}
 We note that the estimates of the
type \eqref{eq:invariance} for problems with time independent
potentials are well-known. They have been proved in the paper by
Hunziker \cite{Hu}. In the time-dependent case the argument is
essentially the same. More precisely, define the functions
$$
\Phi_{j,|\gamma|}(t):=\sum_{j'=0}^{j}\sum_{|\gamma'|=0}^{|\gamma|}
\| \la x\ra^{j'} \pr_{x}^{\gamma'} U(t) g\|_{L^{2}}
$$
for any index $j\ge 0$ and any multi-index $\ga$. Using equation
\eqref{eq:2pot} we obtain that
$$
\frac d{dt} \| \la x\ra^{j} \pr_{x}^{\ga} U(t) g\|_{L^{2}}^{2} = i
(-1)^{|\gamma|}\la \big [\Laplace - V(t,x), \pr_{x}^{\ga}\la
x\ra^{2j}\pr_{x}^{\ga}\big ] U(t)g, U(t)g\ra.
$$
Computing the commutator we obtain the recurrence relation
\begin{align*}
\Phi_{j,|\gamma|}(t)\les& \Phi_{j,|\gamma|}(0) +\la t\ra^{2}
\sum_{|\gamma'|\le 2|\gamma|} \Big \|\frac{\la x\ra}{\la t\ra}
\pr_{x}^{\gamma'} V \Big\|_{L^{\infty}_{t,x}}
 \sup_{0\le \tau \le t} \Phi_{j-1,|\gamma|+1}(\tau)\le\\
&C(V) \bigg (\sum_{k=0}^{j-1} \la
t\ra^{2k}\Phi_{j-k,|\gamma|+k}(0) + \la t\ra^{2j}
\Phi_{0,|\gamma|+j}(\tau)\bigg ),
\end{align*}
where $C(V)$ is a constant depending on
\begin{equation}
\label{eq:potV} \sum_{|\gamma'|\le 2(|\gamma|+j-1)}
\Big\|\frac{\la x\ra}{\la t\ra} \pr_{x}^{\gamma'} V
\Big\|_{L^{\infty}_{t,x}}
\end{equation}
In addition, differentiating the equation \eqref{eq:2pot}
$|\gamma|+j$ times with respect to $x$ and using the standard
$L^{2}$ estimate, we have
$$
\Phi_{0,|\gamma|+j}(\tau)\le C(V) (1+|\tau|^{|\gamma|+j})
\Phi_{0,|\gamma|+j}(0)
$$
Therefore,
$$
\Phi_{j,|\gamma|}(t) \le C(V) (1+|t|)^{3j+|\gamma|}
\Phi_{j,|\gamma|+j}(0).
$$
Now setting $j=j_0 $ and $|\gamma| =\kappa$ we obtain the desired
estimate \eqref{eq:invariance} with $k=j_0 $. Observe that the
assumption  $V_1,V_2 \in C_0^{n+2\kappa+2}(\R^n)$ controls the
constant $C(V)$ in \eqref{eq:potV} for the potential $V(t,x) =
V_{1}(x) + V_{2}(x-t\vec e_{1})$.
\end{proof}

\bibliographystyle{amsplain}

\begin{thebibliography}{99}

\bibitem[Ag]{Ag} Agmon, S. {\em Lectures on exponential decay of
solutions of second-order elliptic equations: bounds on
eigenfunctions of $N$-body Schr\"odinger operators.} Mathematical
Notes, 29. Princeton University Press, Princeton, NJ; University of
Tokyo Press, Tokyo, 1982.


\bibitem[BK]{BK} Ben-Artzi, M., Klainerman, S.,
{\em Decay and regularity for the  Schr\"odinger
 equation. }  J. Anal. Math. 58 (1992), 25--37.

\bibitem[BL]{BL}  Bergh, J., L\"ofstr\"om, J., {\em Interpolation
spaces: an introduction} Berlin, New York; Springer-Verlag, 1976.

\bibitem[Bo]{Bo} Bourgain, J., {\em On long-time behaviour or solutions of linear Schr\"odinger
equations with smooth time-dependent potential} preprint,2002.

\bibitem[En]{En} Enss, V. {\em Geometric Methods in Spectral and Scattering theory of Schr\"odinger operators
} in \textit{Rigorous Atomic and Molecular Physics} ed. by G.
Velo, A.S.Wightman (Plenum, New York 1981) pp. 7-69

\bibitem[GS]{GS} Goldberg, M.,  Schlag, W. {\em Dispersive estimate for the Schr\"odinger
operators in dimensions one and three.} priprint (2003).


\bibitem[Gr]{Gr} Graf, G. M. {\em Phase Space Analysis of the Charge
transfer Model.}   Helv.\ Physica Acta 63 (1990), 107--138.

\bibitem[GV]{GV} Ginibre, J., Velo, G. {\em Generalized Strichartz
Inequalities for the Wave Equation.}  J.\ F.\ Analysis 133 (1995),
50--68.

\bibitem[H]{Hu} Hunziker, W. {\em On the space-time behavior of
Schr\"odinger wavefunctions.} J. Math. Phys. 7 (1966), no. 2, 300--304.

\bibitem[J]{J} Jensen, A. {\em  Spectral properties  of Schr\"odinger operators and time-decay
of the wave functions results in $L^2(\R^m)$, $m \geq 5$.} Duke Math. J. 47
1980, pp 57-80.

\bibitem[JK]{JK} Jensen, A., Kato, T. {\em  Spectral properties  of Schrödinger operators and time-decay of the
wave functions.} Duke Math. J. 46 1979, pp 583-611

\bibitem[JSS]{JSS} Journ\'e, J.-L., Soffer, A., Sogge, C. D. {\em
Decay estimates for Schr\"odinger operators.}
Comm.\ Pure Appl.\ Math.\ 44  (1991),  no.\ 5, 573--604.

\bibitem[KT]{KT} Keel, M., Tao, T. {\em Endpoint Strichartz
Estimates.}  Amer.\ J.\ Math 120 (1998), 955--980.


\bibitem[KY]{KY} Kato, T., Yajima, K.
{\em Some examples of smooth operators and the associated smoothing effect. }
Rev. Math. Phys. 1 (1989), no. 4, 481--496


\bibitem[R]{R} Rauch, J. {\em
Local decay of scattering solutions to Schr\"odinger's equation.}
Comm.\ Math.\ Phys.~61 (1978),
149--168.

\bibitem[RS4]{RS4}  Reed, M., Simon, B. {\em Methods of modern mathematical
physics. IV.} Academic Press [Harcourt Brace Jovanovich, Publishers],
New York-London, 1979.

\bibitem[RS]{RS} Rodnianski, I., Schlag, W. {\em Time decay of solutions
of Schr\"odinger equations with rough and time dependent
potentials.} preprint (2001) to appear in Invent. Math.

\bibitem[RSS1]{RSS1} Rodnianski, I., Schlag, W., Soffer, A.
{\em Dispersive Analysis of Charge Transfer Models.} preprint
(2002), submitted to CPAM.

\bibitem[RSS2]{RSS2} Rodnianski, I., Schlag, W., Soffer, A.
{\em Asymptotic stability of N-soliton states of NLS.} preprint
(2003), submitted to CPAM.

\bibitem[So]{So} Soffer, A. {\em On the many-body problem in quantum mechanics.}
Ast\'erisque No. 207 (1992), 6, 109--152.

\bibitem[Vai]{V} Va\u\i nberg, B. R.  {\em The short-wave asymptotic
behavior of the solutions of stationary problems,
and the asymptotic behavior as $t\rightarrow \infty $ of the
solutions of nonstationary problems.}  (Russian)
Uspehi Mat.\ Nauk  30  (1975), no.~2 (182), 3--55.

\bibitem[Wu]{Wu} W\"uller, U. {\em Geometric Methods in Scattering
Theory of the Charge Transfer Model.}
Duke Math J. 62, no.~2 (1991), 273--313.

\bibitem[Ya1]{Ya1} Yajima, K. {\em The $W\sp {k,p}$-continuity of
wave operators for Schr\"odinger operators.}
J.\ Math.\ Soc.\ Japan  47  (1995),  no.~3, 551--581.

\bibitem[Ya2]{Ya2} Yajima, K. {\em A multichannel scattering
theory for some time dependent Hamiltonians, charge transfer
problem.}
Comm.\ Math.\ Phys.\ 75  (1980), no.~2, 153--178.

\bibitem[Zi]{Z} Zielinski, L. {\em Asymptotic completeness
for multiparticle dispersive charge transfer models.} J. Funct. Anal.
150 (1997), no. 2, 453--470.

\end{thebibliography}

\noindent \textsc{Kaihua Cai: Division of Astronomy, Mathematics,
and Physics,
253-37 Caltech, Pasadena, C.A. 91125, U.S.A.}\\
{\em email: }\textsf{\bf kaihua@its.caltech.edu}

\end{document}